\documentclass{article}

\PassOptionsToPackage{numbers, sort&compress}{natbib}
\usepackage[preprint]{neurips_2026}
\workshoptitle{Foundation Models for the Brain and Body}

\usepackage{comment}
\usepackage[utf8]{inputenc}
\usepackage[T1]{fontenc}
\usepackage{hyperref}
\usepackage{url}
\usepackage{booktabs}
\usepackage{amsmath}
\usepackage{amsfonts}
\usepackage{nicefrac}
\usepackage{microtype}
\usepackage{xcolor}
\usepackage{graphicx}
\usepackage{longtable}
\usepackage{float}
\usepackage{array}
\usepackage[skip=4pt]{caption}
\usepackage{tikz}
\usetikzlibrary{shapes.geometric,arrows.meta,positioning,fit,backgrounds}

\usepackage{enumitem}
\setlist[itemize]{topsep=2pt,itemsep=1pt,parsep=0pt,partopsep=0pt,leftmargin=1.15em}
\setlist[itemize,2]{topsep=1pt,itemsep=0pt,parsep=0pt,leftmargin=1.1em}

\usepackage{adjustbox}
\usepackage{colortbl}
\usepackage{upquote}
\newcommand{\critrule}{\arrayrulecolor{black!22}\specialrule{0.3pt}{2.5pt}{2.5pt}\arrayrulecolor{black}}
\newcommand{\qsep}{\arrayrulecolor{black!20}\specialrule{0.3pt}{3pt}{3pt}\arrayrulecolor{black}}
\newcommand{\passrule}{\arrayrulecolor{black!70}\specialrule{0.9pt}{5pt}{4pt}\arrayrulecolor{black}}
\let\oldtabular\tabular
\let\endoldtabular\endtabular
\renewenvironment{tabular}%
  {\begin{adjustbox}{max width=\linewidth}\oldtabular}%
  {\endoldtabular\end{adjustbox}}

\title{Foundation Models as Modular Tools \\ for fMRI Analysis}

\author{%
  Joseph Kilgore \\
  Brookhaven National Laboratory \\
  \And
  Noemia Mahmud \\
  Brookhaven National Laboratory \\
  \AND
  Shinjae Yoo \\
  Brookhaven National Laboratory \\
  \And
  David Keetae Park \\
  Brookhaven National Laboratory \\
  \texttt{dpark1@bnl.gov} \\
}

\begin{document}
\maketitle

\begin{abstract}
Foundation models (FMs) for the brain are rapidly growing in number and each captures a different aspect of the same data. A single scientific question often needs several of them at once, alongside classical analysis and the literature, but no established method combines them, because each model defines its own inputs, outputs, and spatial conventions. Representation alignment, the usual route from an FM into a language model, binds one FM to one LLM and requires retraining for every model added. We take a modular route: each FM becomes an independent tool server that a fixed language agent calls next to classical and literature tools, with no retraining. We start with fMRI, where public FMs already disagree on how to name a brain location and open task data give a classical baseline, and serve two FMs at opposite ends of that spectrum, BrainLM (atlas parcels) and Omni-fMRI (atlas-free voxels). A 76-question benchmark, a 2 by 2 tool ablation, and an independent LLM judge yield 3{,}040 judged traces. In this first study, modular FM access raises holistic answer quality by $+0.70$ on a 1 to 5 rubric (95\% CI $[+0.44, +0.97]$, $n=34$). The gain lands in how the agent selects, parameterizes, and grounds tool calls, and peaks where the agent must cross FM vocabularies. The two FMs win on different questions, yet mounting both matches the better one, and the voxel interface fails 7.8 times more often by context overflow. Modular connection works. Orchestrating what it connects, across EEG and MRI models, is the work ahead.
\end{abstract}

\section{Introduction}
\label{sec:intro}

Foundation models (FMs) for the brain are gaining substantial attention in EEG \cite{yang2023biot,jiang2024labram,wang2024eegpt,wang2025cbramod}, fMRI \cite{kim2023swift,caro2024brainlm,dong2024brainjepa,wang2026omnifmri}, and structural MRI \cite{ma2024medsam,cox2024brainsegfounder}. These models do not interoperate, since each uses its own spatial vocabulary and reports its own benchmark, with no mapping between them. Applying a second model to the same scan therefore requires learning its interface and translating every region of interest between the two coordinate conventions, so most analyses rely on a single model by convention.

Fragmentation also carries a scientific cost, because the differences between models are informative. A model that summarizes a scan over a named anatomical atlas and one that operates in the raw measurement space \cite{caro2024brainlm,wang2026omnifmri} describe the same data at different spatial resolution and report different quantities, and they degrade on different scans. Disagreement between them therefore reflects a property of the scan rather than noise, and no single model reports it. Obtaining such evidence requires querying several models within one analysis, which current workflows do not support.

The standard way to give a language model access to an FM is representation alignment \cite{alayrac2022flamingo,li2023blip2}. A learned projector maps FM embeddings into the LLM token space, and the two train jointly on paired data, as in LLaVA for vision \cite{liu2023llava} and NeuroLM for EEG \cite{jiang2025neurolm}. Alignment gives the LLM one model's semantics, but it binds the two together, so the aligned system sees the data only through that FM, and adding a second model requires new paired data and a further training run.

We take a modular alternative in which each FM runs as an independent tool server under a standard schema, the Model Context Protocol (MCP) \cite{anthropic2024mcp}, alongside classical analysis tools and a literature server, and a fixed language agent calls whichever tools it judges relevant in a reason-and-act loop \cite{yao2023react,schick2023toolformer}. The design requires no training and no alignment, so adding a model means adding a server. Agents that orchestrate other models already exist for general-purpose tasks \cite{shen2023hugginggpt,qin2024toolllm}, but scientific FMs raise two questions those systems do not face, namely whether an agent can invoke a brain FM it has never seen from tool descriptions alone, and what an interface costs when incompatible vocabularies share one context window.

Both questions concern the interface rather than the representation. An FM reaches the final answer only through the calls an agent can make and the tokens those calls consume, so two models that encode a scan equally well can differ sharply in how usable they are. fMRI makes this concrete, because its public models disagree on how to name a location in the brain. BrainLM takes named A424 cortical parcels \cite{caro2024brainlm}, Omni-fMRI takes raw voxel coordinates \cite{wang2026omnifmri}, the classical toolchain uses MNI seed names, and no lookup table relates the three. fMRI also offers open task datasets and a mature classical analysis suite, which give us a baseline that uses no FM at all. We serve both models as separate MCP servers alongside 54 classical tools and a literature server (Appendix~\ref{app:servers}), all reading the same scans from AOMIC PIOP1 across seven task stages \cite{snoek2021aomic}.
 
We measure the design by ablation, holding the agent, dataset, 76 questions (Appendix~\ref{app:questions}), and judge fixed and varying only which FM servers we mount. This gives four conditions and 3{,}040 judged traces, each scored on ten criteria (Appendix~\ref{app:rubric}) by an independent judge from a different model family \cite{zheng2023judging}. The ablation tests modularity directly because the agent trained on neither model and met each one through its tool descriptions at run time.

\paragraph{Contributions.}
(1) \textbf{A modular FM-as-a-tool system for neuroimaging, and the benchmark that tests it.} FMs with incompatible addressing schemes attach to one agent that never trains on them, and 76 designed questions with a ten-criterion judge measure what that access adds to an analysis.
(2) \textbf{Modular access improves analysis quality, and the improvement concentrates at the tool interface.} Against the same agent with classical tools only, mounting the FMs raises judged answer quality by $+0.70$ on a 1 to 5 rubric (95\% CI $[+0.44, +0.97]$), and each model helps on its own. With an FM available, the agent selects and configures tools better (tool selection $+1.09$) but reads the returned values no better, and the gain is largest on questions that require translating a brain location from one model's naming scheme to another ($+1.04$).
(3) \textbf{Two failure modes in multi-FM use.} The two models answer different questions well, but an agent with both scores the same as an agent with BrainLM alone ($+0.04$), so the gain stops at whatever one model already provides. Separately, the voxel-addressed model ends 7.8 times as many runs without a gradeable answer ($4.84\%$ versus $0.62\%$), almost always because listing coordinates fills the agent's context window.

\section{A modular FM-as-a-tool system}
\label{sec:system}

\paragraph{Registry of tool servers.}
The agent is a fixed LLM (\texttt{mistral-24b-fp8}, 131{,}072-token context) in an iterative tool-calling loop over four MCP servers: \texttt{piop1} (54 classical tools for time-series, GLM contrasts, and graph metrics), \texttt{pubmed} (8 literature tools), \texttt{brainlm} (10 tools for reconstruction error, forecasting, and state similarity), and \texttt{omni\_fmri} (11 for patch embeddings and error maps). Each server registers its tools with a text description and a discovery call, and nothing else connects them, so the agent learns a vocabulary at run time (Appendix~\ref{app:servers}).

\paragraph{Three ways to name a brain location.}
The three servers that hold data each name brain locations differently, and no lookup table converts between them. \texttt{piop1} takes an MNI seed name used only for orientation (\texttt{frontal\_inf\_l}), \texttt{brainlm} a named A424 cortical parcel (\texttt{region="L\_44"}), and \texttt{omni\_fmri} a raw voxel coordinate (\texttt{x=48, y=48, z=48}). We call these three conventions \textit{addressing schemes}. All three servers read the same scans, AOMIC PIOP1 (ds002785) \cite{snoek2021aomic}, across seven task stages.

\paragraph{The modularity test.}
\texttt{piop1} and \texttt{pubmed} stay mounted in every condition, and the two FM servers form a $2\times2$ ablation over \texttt{classical}, \texttt{brainlm\_only}, \texttt{omni\_only}, and \texttt{full}. Tool exposure is the only independent variable, so the contrasts characterize the FMs, but not the agent.

\section{Benchmark and measurement}
\label{sec:benchmark}

\paragraph{Question set.}
We wrote 76 questions in ten families along three axes. Axis A, \emph{single-FM workflow}, asks whether an FM earns a place inside a realistic multi-step analysis, through XL extra-long chains (8), ES essential (3, the FM supplies a quantity no classical tool returns), EN enhancing (3, a complete classical answer already exists), and EB efficiency-bounded (3, a fixed call budget). Axis B, \emph{interface crossing}, targets fragmentation directly, through VC vocabulary crossing (14) and CB calibrated bridge (3), which force the agent to move a region of interest across mismatched vocabularies, and SX silent versus explicit (6), which varies whether the prompt warns about the mismatch. Axis C, \emph{multi-FM synthesis}, puts the same question to both FMs and asks whether they agree, through SY (24). LC literature (10) and ZT zero-tool trap (2) serve as controls, and Appendix~\ref{app:questions} lists every prompt.

\paragraph{Scoping.}
CORE, our primary set, holds the 34 questions of VC, XL, ES, EN, CB, and EB. We exclude four families for reasons that hold in every condition and so cannot favour any arm (Table~\ref{tab:scope}). SY and SX carry an A/B wording contrast worth $+0.65$ to $+0.94$ even under \texttt{classical}, LC suffers condition-independent harness failures, and ZT sits at ceiling everywhere. We did not choose the scope blind to results, so we report every contrast on both CORE and all 76 questions and treat Axis C as exploratory. Exclusion did not select for the effect, since the excluded LC family has a negative difference and the excluded SX family the largest positive one. One asymmetry limits any head-to-head comparison, namely that 12 prompts name BrainLM's tools and only 1 names Omni-fMRI's.

\paragraph{Runs, judge, and statistics.}
Each question ran under all four conditions with ten seeds, for 3{,}040 traces, of which 53 ($1.7\%$) returned no gradeable output. We exclude those from means and treat their distribution as a measurement in Section~\ref{sec:interface}. A judge from a different model family (\texttt{llama3.3-70b-fp8}) scored every trace in three passes on ten criteria from 1 to 5 (Appendix~\ref{app:rubric}), enforcing primitive discipline, so an answer that reads a reconstruction error as a diagnosis loses points at any value. For each question we average \texttt{holistic\_score} over its ten seeds within a condition and take the paired difference against \texttt{classical}, so the per-question differences form the sample. Intervals are BCa bootstraps over questions, $p$-values come from sign-permutation tests, per-criterion tests carry Benjamini--Hochberg correction, and the two-FM comparison uses two one-sided tests against a $\pm0.25$ margin fixed in advance as half the primary effect. A mixed model with question as a random effect reproduces the estimates (Appendix~\ref{app:stats}).

\begin{table}[H]
\centering
\small
\caption{Every question family, its mean full-minus-classical difference, and whether it is in CORE. Families are excluded for reasons that hold equally in every condition, not for being unfavourable: note that the excluded LC family has a \emph{negative} difference and the excluded SX family has the largest positive one.}
\label{tab:scope}
\begin{tabular}{lrrll}
\toprule
Family & $n$ & $\Delta$ & In CORE? & Reason \\
\midrule
VC & 14 & +1.04 & CORE & primary \\
SY & 24 & +0.29 & excluded & A/B wording effect, present in every condition including classical \\
LC & 10 & -0.18 & excluded & harness failures independent of condition (parser error; timeouts) \\
XL & 8 & +0.63 & CORE & primary \\
SX & 6 & +1.70 & excluded & same A/B wording effect as SY \\
CB & 3 & +0.09 & CORE & primary \\
ES & 3 & +0.64 & CORE & primary \\
EN & 3 & +0.24 & CORE & primary \\
EB & 3 & +0.38 & CORE & primary \\
ZT & 2 & +0.05 & excluded & at ceiling in every condition (4.84--5.00) \\
\bottomrule
\end{tabular}
\end{table}

\section{Results}
\label{sec:results}

\subsection{Modular FM access improves analysis quality, at the tool interface}
\label{sec:headline}
Mounting both FMs raises the holistic score by $+0.70$ over \texttt{classical} (95\% CI $[+0.44, +0.97]$, $n=34$), and each model helps alone, $+0.66$ for BrainLM and $+0.44$ for Omni-fMRI (Table~\ref{tab:primary}). Eighteen of the 34 CORE questions improve by more than 0.5, 14 stay flat, and 2 decline, and on all 76 questions the effect is $+0.52$ ($[+0.32, +0.72]$). A worst-case imputation of the ungradeable runs leaves $+0.67$ (Appendix~\ref{app:tables}).

\begin{figure}[t]
  \centering
  \includegraphics[width=\linewidth]{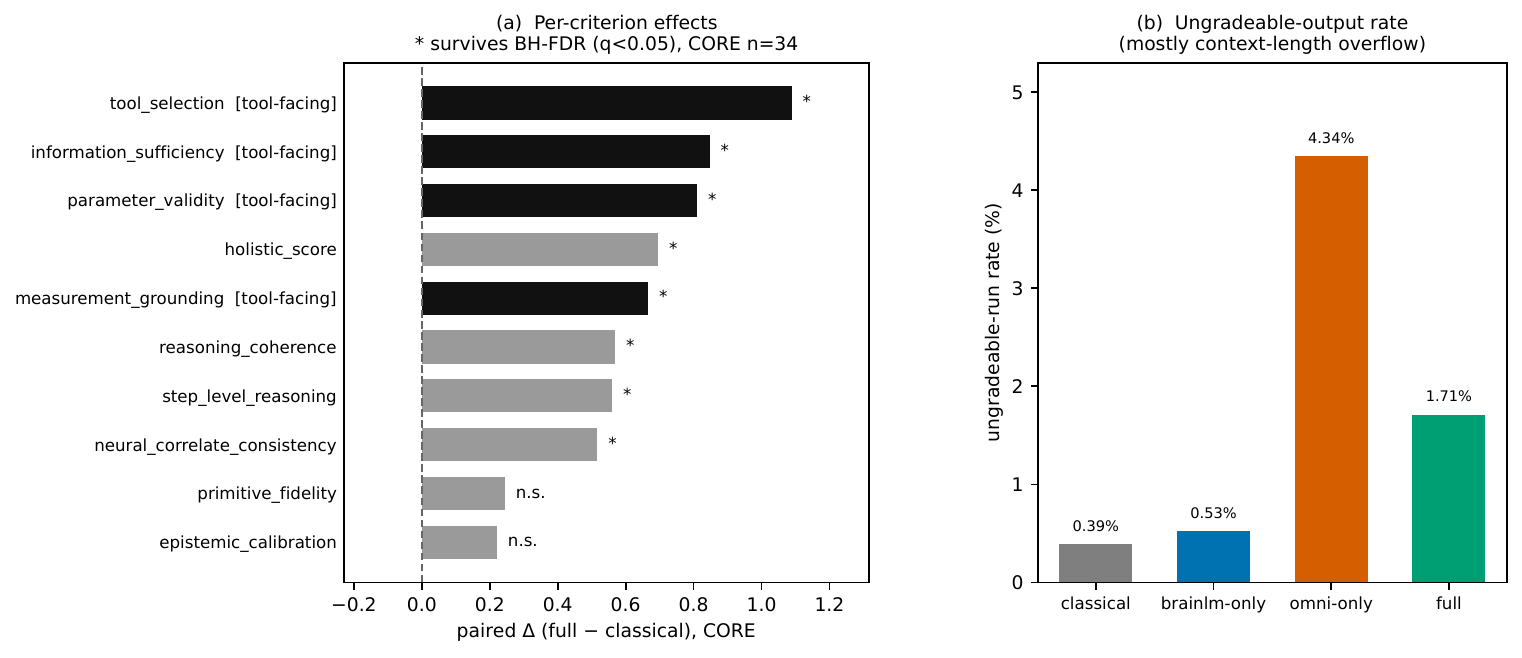}

  \caption{What modular FM access changes, and what an addressing scheme costs.
  (a) Per-criterion differences, marked for BH-FDR significance.
  (b) Ungradeable-run rate by condition over all 76 questions. Almost every \texttt{omni\_only} failure is a context overflow. Section~\ref{sec:interface} quotes the higher rate on the 64 questions that do not name BrainLM's tools.}
  \label{fig:headline}
\end{figure}

The gain is not spread evenly across the rubric (Figure~\ref{fig:headline}a). The four criteria describing how the agent uses tools (\texttt{tool-facing}) move most and all survive correction, led by tool selection $+1.09$, information sufficiency $+0.85$, parameter validity $+0.81$, and measurement grounding $+0.67$, while the two describing how it reads back a value, primitive fidelity ($+0.24$) and epistemic calibration ($+0.22$), are not significant. An untrained agent therefore discovers what an FM server offers, selects it when it fits, parameterizes it in the server's vocabulary, and grounds claims in what it returns, but it reads the returned number no better, which is where alignment would enter.

\begin{table}[H]
\centering
\small
\caption{Effect of foundation-model access on holistic answer quality.}
\label{tab:primary}
\begin{tabular}{lrrrrr}
\toprule
Contrast & $n$ & $\Delta$ & 95\% CI & $p$ & up / flat / down \\
\midrule
BrainLM only $-$ classical & 34 & +0.66 & [+0.42, +0.92] & $<$0.001 & 18 / 16 / 0 \\
Omni only $-$ classical & 34 & +0.44 & [+0.22, +0.68] & $<$0.001 & 16 / 15 / 3 \\
Both FMs $-$ classical & 34 & +0.70 & [+0.44, +0.97] & $<$0.001 & 18 / 14 / 2 \\
Both FMs $-$ classical \emph{(all 76 questions)} & 76 & +0.52 & [+0.32, +0.72] & $<$0.001 & 35 / 32 / 9 \\
\bottomrule
\end{tabular}
\end{table}

\subsection{The gain peaks where fragmentation bites}
\label{sec:crossing}
Grouping the families by axis locates the gain (Table~\ref{tab:scope}). Axis A, single-FM workflows, improves by $+0.52$, and its ordering matches the design, since ES questions, where the FM supplies a quantity classical tools cannot, gain more than EN questions, where a complete classical answer already exists. Axis B, interface crossing, improves by $+1.09$, VC alone by $+1.04$, and the excluded SX family, which varies only whether the prompt warns of the mismatch, by $+1.70$. VC-07 shows the mechanism in one question. It asks the agent to call BrainLM with a region name copied from \texttt{piop1}'s vocabulary, report the failure, and recover through BrainLM's discovery call (Appendix~\ref{app:traces}). Without the server mounted the agent cannot reach the error and narrates an outcome it never observed, while with it mounted the call fails for the intended reason and the agent recovers to a valid parcel and a real measurement. Modular access does not remove fragmentation. Rather, it gives the agent the discovery calls it needs to work across it.

\subsection{Two FMs: complementary views that co-mounting does not harvest}
\label{sec:twofm}
Mounting a second FM next to the first yields no detectable benefit on the primary set. \texttt{full} minus \texttt{brainlm\_only} is $+0.04$ (90\% CI $[-0.12, +0.19]$), inside the pre-specified $\pm0.25$ margin ($p_{\mathrm{TOST}}=0.011$), while the comparison against \texttt{omni\_only} is $+0.25$ ($[+0.06, +0.44]$) and stays inconclusive (Appendix~\ref{app:tables}). Neither model outperforms the other once we control for which model a question names, since on the 20 CORE questions naming both equally BrainLM leads by only $+0.15$ ($[-0.17, +0.51]$) (Appendix~\ref{app:directedness}).

Equivalence does not mean the second model carries no information. The two FMs win on different questions, BrainLM leading by 0.5 or more on 19 of the 76 and trailing by as much on 13. Picking the better model per question would score 3.77 against 3.62 for both mounted, an upper bound because selection and scoring share seeds. Axis C, which puts the same question to both models, is the one place where \texttt{full} is the best arm (3.10 against 2.94 and 3.01), though the family is confounded and we state this as a direction. Complementarity therefore exists at the question level, an agent given both servers does not collect it, and orchestration is the problem that modular connection exposes rather than solves.

\subsection{An interface has a cost: reliability, not accuracy}
\label{sec:interface}
The clearest difference between the two models is whether the agent returns an answer at all. On the 64 questions that do not name BrainLM's tools, \texttt{omni\_only} fails to produce a gradeable answer on $4.84\%$ of runs against $0.62\%$ for \texttt{brainlm\_only}, a factor of 7.8 (15 of 64 questions show at least one failure against 2, sign test $p=0.004$). The failures are of one kind, since 29 of the 31 \texttt{omni\_only} failures are context overflows at the 131{,}072-token limit while none of \texttt{brainlm\_only}'s 4 overflow, and only 2 fall on BrainLM-directed questions, so directedness does not explain the gap (Figure~\ref{fig:headline}b). Enumerating voxel coordinates consumes context that named parcels do not, and the agent exhausts its budget before it exhausts the analysis.

This is an interface property, and thus should be considered when architecting further FMs. A voxel-addressed model that also exposed an aggregation call, such as a mean over an atlas region, would pay far less of this cost. Under a fixed context window, the addressing scheme an FM exposes impacts the agents access and usage.

\section{Discussion and outlook}
\label{sec:discussion}

A scientific question often needs more than one foundation model, but the models do not compose, and aligning one to a language model shuts out the others. We asked whether serving FMs as ordinary tools lets one agent use several. Over 3{,}040 judged traces, access to two fMRI FMs raised answer quality by $+0.70$ on a 1 to 5 rubric. The improvement appeared in how the agent chose and configured its calls, and it was largest when a question crossed the two models' naming conventions. Three interface properties drove that result, and none requires changing the model. A discovery call lets the agent learn the model's vocabulary at run time, a bounded output size keeps one call from filling the context that later calls need, and outputs stated as measurements keep the agent to what a value supports. An FM that provides all three can work alongside other models as soon as it is released.
 
We used one LLM as the agent and one as the judge, and a different pair may give different numbers. Our questions also mention BrainLM's tools far more often than Omni-fMRI's, so the comparison between the two models is uneven and we do not rank them. The main open problem is orchestration. Giving the agent both models scored the same as giving it one, so the next step is a policy that picks the right model for each question and treats disagreement between models as a signal. Nothing here is specific to fMRI. EEG, structural, and functional models can serve one agent together, and the same design extends to domains where a question spans instruments, simulations, and literature. There too, an FM's interface will matter as much as its representation.

\section*{Acknowledgments}
This work was supported by the U.S. Department of Energy (DOE), Office of Science (SC), Advanced Scientific Computing Research program under award DE-SC-0012704 and used resources of the National Energy Research Scientific Computing Center, a DOE Office of Science User Facility using NERSC award NERSC DDR-ERCAP0030592.

\small
\bibliographystyle{plainnat}
\bibliography{refs}

@inproceedings{caro2024brainlm,
  title     = {{BrainLM}: A foundation model for brain activity recordings},
  author    = {Ortega Caro, Josue and de O. Fonseca, Antonio H. and
               Rizvi, Syed A. and Rosati, Matteo and Cross, James L. and
               Dhodapkar, Rahul M. and van Dijk, David},
  booktitle = {International Conference on Learning Representations (ICLR)},
  year      = {2024},
  note      = {carried from v2}
}

@inproceedings{wang2026omnifmri,
  title     = {{Omni-fMRI}: A universal atlas-free {fMRI} foundation model},
  author    = {Wang, M. and Ye, W. and Xia, J. and Zhang, J. and Pan, X. and
               Xu, M. and Deng, H. and Wen, H. and Liu, Q.},
  booktitle = {International Conference on Machine Learning (ICML)},
  year      = {2026},
  note      = {carried from v2}
}

@inproceedings{kim2023swift,
  title     = {{SwiFT}: Swin 4{D} {fMRI} transformer},
  author    = {Kim, Peter Y. and Kwon, Junbeom and Joo, Sunghwan and
               Bae, Sangyoon and Lee, Donggyu and Jung, Yoonho and
               Yoo, Shinjae and Cha, Jiook and Moon, Taesup},
  booktitle = {Advances in Neural Information Processing Systems (NeurIPS)},
  year      = {2023},
  note      = {VERIFY author list and page numbers}
}

@inproceedings{dong2024brainjepa,
  title     = {{Brain-JEPA}: Brain dynamics foundation model with gradient
               positioning and spatiotemporal masking},
  author    = {Dong, Zijian and Li, Ruilin and Wu, Yilei and
               Nguyen, Thuan Tinh and Chong, Joanna Su Xian and
               Ji, Fang and Tong, Nathanael Ren Jie and
               Chen, Christopher Li Hsian and Zhou, Juan Helen},
  booktitle = {Advances in Neural Information Processing Systems (NeurIPS)},
  year      = {2024},
  note      = {VERIFY author list}
}

@inproceedings{yang2023biot,
  title     = {{BIOT}: Biosignal transformer for cross-data learning in the wild},
  author    = {Yang, Chaoqi and Westover, M. Brandon and Sun, Jimeng},
  booktitle = {Advances in Neural Information Processing Systems (NeurIPS)},
  year      = {2023},
  note      = {VERIFY page numbers}
}

@inproceedings{jiang2024labram,
  title     = {Large brain model for learning generic representations with
               tremendous {EEG} data in {BCI}},
  author    = {Jiang, Wei-Bang and Zhao, Li-Ming and Lu, Bao-Liang},
  booktitle = {International Conference on Learning Representations (ICLR)},
  year      = {2024},
  note      = {VERIFY}
}

@inproceedings{wang2024eegpt,
  title     = {{EEGPT}: Pretrained transformer for universal and reliable
               representation of {EEG} signals},
  author    = {Wang, Guangyu and Liu, Wenchao and He, Yuhong and
               Xu, Cong and Ma, Lin and Li, Haifeng},
  booktitle = {Advances in Neural Information Processing Systems (NeurIPS)},
  year      = {2024},
  note      = {VERIFY author list}
}

@inproceedings{wang2025cbramod,
  title     = {{CBraMod}: A criss-cross brain foundation model for {EEG} decoding},
  author    = {Wang, Jiquan and Zhao, Sha and Luo, Zhiling and Zhou, Yangxuan
               and Jiang, Haiteng and Li, Shijian and Li, Tao and Pan, Gang},
  booktitle = {International Conference on Learning Representations (ICLR)},
  year      = {2025},
  note      = {VERIFY author list}
}

@article{ma2024medsam,
  title   = {Segment anything in medical images},
  author  = {Ma, Jun and He, Yuting and Li, Feifei and Han, Lin and
             You, Chenyu and Wang, Bo},
  journal = {Nature Communications},
  volume  = {15},
  pages   = {654},
  year    = {2024},
  note    = {VERIFY}
}

@article{cox2024brainsegfounder,
  title   = {{BrainSegFounder}: Towards 3{D} foundation models for
             neuroimage segmentation},
  author  = {Cox, Joseph and Liu, Peng and Stolte, Skylar E. and
             Yang, Yunchao and Liu, Kang and See, Kyle B. and
             Ju, Huiwen and Fang, Ruogu},
  journal = {Medical Image Analysis},
  year    = {2024},
  note    = {VERIFY volume and pages}
}

@inproceedings{alayrac2022flamingo,
  title     = {Flamingo: A visual language model for few-shot learning},
  author    = {Alayrac, Jean-Baptiste and Donahue, Jeff and Luc, Pauline and
               Miech, Antoine and Barr, Iain and Hasson, Yana and
               Lenc, Karel and Mensch, Arthur and Millican, Katherine and
               Reynolds, Malcolm and others},
  booktitle = {Advances in Neural Information Processing Systems (NeurIPS)},
  year      = {2022},
  note      = {VERIFY page numbers}
}

@inproceedings{li2023blip2,
  title     = {{BLIP-2}: Bootstrapping language-image pre-training with frozen
               image encoders and large language models},
  author    = {Li, Junnan and Li, Dongxu and Savarese, Silvio and Hoi, Steven},
  booktitle = {International Conference on Machine Learning (ICML)},
  year      = {2023},
  note      = {VERIFY page numbers}
}

@inproceedings{liu2023llava,
  title     = {Visual instruction tuning},
  author    = {Liu, Haotian and Li, Chunyuan and Wu, Qingyang and Lee, Yong Jae},
  booktitle = {Advances in Neural Information Processing Systems (NeurIPS)},
  year      = {2023},
  note      = {VERIFY page numbers}
}

@inproceedings{jiang2025neurolm,
  title     = {{NeuroLM}: A universal multi-task foundation model for bridging
               the gap between language and {EEG} signals},
  author    = {Jiang, Wei-Bang and Wang, Yansen and Lu, Bao-Liang and Li, Dongsheng},
  booktitle = {International Conference on Learning Representations (ICLR)},
  year      = {2025},
  note      = {VERIFY author list}
}

@inproceedings{yao2023react,
  title     = {{ReAct}: Synergizing reasoning and acting in language models},
  author    = {Yao, Shunyu and Zhao, Jeffrey and Yu, Dian and Du, Nan and
               Shafran, Izhak and Narasimhan, Karthik and Cao, Yuan},
  booktitle = {International Conference on Learning Representations (ICLR)},
  year      = {2023},
  note      = {carried from v2}
}

@inproceedings{schick2023toolformer,
  title     = {Toolformer: Language models can teach themselves to use tools},
  author    = {Schick, Timo and Dwivedi-Yu, Jane and Dess{\`\i}, Roberto and
               Raileanu, Roberta and Lomeli, Maria and Zettlemoyer, Luke and
               Cancedda, Nicola and Scialom, Thomas},
  booktitle = {Advances in Neural Information Processing Systems (NeurIPS)},
  year      = {2023},
  note      = {carried from v2}
}

@inproceedings{shen2023hugginggpt,
  title     = {{HuggingGPT}: Solving {AI} tasks with {ChatGPT} and its friends
               in {Hugging Face}},
  author    = {Shen, Yongliang and Song, Kaitao and Tan, Xu and Li, Dongsheng
               and Lu, Weiming and Zhuang, Yueting},
  booktitle = {Advances in Neural Information Processing Systems (NeurIPS)},
  year      = {2023},
  note      = {VERIFY page numbers}
}

@inproceedings{qin2024toolllm,
  title     = {{ToolLLM}: Facilitating large language models to master
               16000+ real-world {APIs}},
  author    = {Qin, Yujia and Liang, Shihao and Ye, Yining and Zhu, Kunlun and
               Yan, Lan and Lu, Yaxi and Lin, Yankai and Cong, Xin and
               Tang, Xiangru and Qian, Bill and others},
  booktitle = {International Conference on Learning Representations (ICLR)},
  year      = {2024},
  note      = {VERIFY author list}
}

@inproceedings{zheng2023judging,
  title     = {Judging {LLM-as-a-judge} with {MT-Bench} and {Chatbot Arena}},
  author    = {Zheng, Lianmin and Chiang, Wei-Lin and Sheng, Ying and
               Zhuang, Siyuan and Wu, Zhanghao and Zhuang, Yonghao and
               Lin, Zi and Li, Zhuohan and Li, Dacheng and Xing, Eric P. and
               Zhang, Hao and Gonzalez, Joseph E. and Stoica, Ion},
  booktitle = {Advances in Neural Information Processing Systems (NeurIPS)},
  year      = {2023},
  note      = {carried from v2}
}

@techreport{anthropic2024mcp,
  title       = {Introducing the {Model Context Protocol}},
  author      = {{Anthropic}},
  institution = {Anthropic},
  year        = {2024},
  url         = {https://www.anthropic.com/news/model-context-protocol},
  note        = {carried from v2}
}

@article{snoek2021aomic,
  title   = {The {Amsterdam Open MRI Collection}, a set of multimodal {MRI}
             datasets for individual difference analyses},
  author  = {Snoek, Lukas and van der Miesen, Maite M. and
             Beemsterboer, Tinka and van der Leij, Andries and
             Eigenhuis, Annemarie and Scholte, H. Steven},
  journal = {Scientific Data},
  volume  = {8},
  pages   = {85},
  year    = {2021},
  note    = {carried from v2}
}
\normalsize

\clearpage
\appendix

\section{Judge rubric}
\label{app:rubric}
Every run was scored by an independent LLM judge (\texttt{llama3.3-70b-fp8})
in three passes over the trace, each with its own system prompt and its own
set of 1--5 integer criteria. The passes are deliberately separated so that a
good final answer reached by a bad trajectory cannot borrow credit from the
tool-use pass, and vice versa: the tool-accuracy prompt states that it is
\emph{not} evaluating the final inference, and the inference prompt that it is
\emph{not} re-evaluating the tool calls. Table~\ref{tab:rubric} gives the ten
criteria with their endpoint anchors; the full prompts, including the
intermediate anchors, are in \texttt{criteria\_judge\_00.csv}.

A constraint common to the first two passes is what we call
\emph{primitive discipline}: NeuroX tools return measurements, not verdicts.
A scalar such as \texttt{reconstruction\_surprise}${}=0.83$ is a measurement of
atypicality relative to a training distribution, not a diagnosis, and an agent
step that treats it as one is scored as a fidelity failure regardless of how
extreme the value is. Valid inference language is explicitly enumerated for
the judge (``consistent with'', ``suggestive of'', ``patterns associated
with'', ``elevated relative to baseline''); categorical diagnostic claims are
epistemic failures at any measurement value.

\begin{table}[H]
\centering
\caption{The ten judge criteria, grouped by evaluation pass, with the 1 and 5
anchors. Each is an integer 1--5. \texttt{holistic\_score}, the headline
outcome used throughout this paper, is emitted by the trajectory pass as an
overall judgment of the run.}
\label{tab:rubric}
\footnotesize
\setlength{\tabcolsep}{4pt}
\begin{tabular}{@{}>{\raggedright\arraybackslash}p{2.5cm}
                 >{\raggedright\arraybackslash}p{5.15cm}
                 >{\raggedright\arraybackslash}p{5.15cm}@{}}
\toprule
criterion & 1 & 5 \\
\midrule
\multicolumn{3}{@{}p{13.3cm}@{}}{\textit{Pass 1 --- tool accuracy} (evaluates the tool-call sequence, not the inference)} \\[2pt]
\texttt{tool\_selection} &
  Entirely wrong tools; cannot address the query &
  Exactly the right tools, right order, no unnecessary calls \\ \critrule
\texttt{parameter\_\allowbreak validity} &
  Fundamentally wrong parameters (nonexistent ROIs, wrong subject or run IDs) &
  All parameters fully correct and well chosen \\ \critrule
\texttt{primitive\_\allowbreak fidelity} &
  Treated a tool output as a diagnostic verdict &
  Treated all outputs as measurements; no unsupported conclusions from any single primitive \\
\passrule
\multicolumn{3}{@{}p{13.3cm}@{}}{\textit{Pass 2 --- inference quality} (evaluates the final claim, not the tool calls)} \\[2pt]
\texttt{measurement\_\allowbreak grounding} &
  Contradicts the measurements, or cites measurements never collected &
  Every claim traces to a specific tool output \\ \critrule
\texttt{neural\_\allowbreak correlate\_\allowbreak consistency} &
  Contradicts well-established neural correlates for the claimed state &
  Accurately reflects established correlates and hedges where the literature is contested \\ \critrule
\texttt{epistemic\_\allowbreak calibration} &
  Categorical claims (``the subject is anxious'') from population-level fMRI &
  Confident where measurements converge, explicitly uncertain where they do not \\
\passrule
\multicolumn{3}{@{}p{13.3cm}@{}}{\textit{Pass 3 --- trajectory reasoning} (evaluates the path, not the destination)} \\[2pt]
\texttt{reasoning\_\allowbreak coherence} &
  Disorganized; tool calls disconnected from each other and from the query &
  Each call explicitly motivated by prior output, moving efficiently toward the query \\ \critrule
\texttt{information\_\allowbreak sufficiency} &
  Concluded without gathering basic supporting evidence &
  Complete coverage of relevant measurement dimensions without redundancy \\ \critrule
\texttt{step\_level\_\allowbreak reasoning} &
  No reasoning between calls; outputs accumulated without interpretation &
  Each measurement explicitly linked to prior evidence and to the query \\ \critrule
\texttt{holistic\_score} &
  \multicolumn{2}{@{}>{\raggedright\arraybackslash}p{10.4cm}@{}}{Overall trajectory judgment, 1--5 (see note below)} \\
\bottomrule
\end{tabular}
\end{table}


The trajectory pass additionally tells the judge what a good trajectory looks
like, and \texttt{holistic\_score} should be read against that description:
appropriate initial scoping; each call motivated by what earlier calls
returned; coverage of several measurement dimensions before inferring; no
repeated or unused calls; stopping once the evidence is sufficient; and
primitive discipline maintained at every intermediate step, not only in the
final answer.

\section{Tool servers}
\label{app:servers}
\begin{table}[H]
\centering
\small
\caption{The four MCP tool servers.}
\label{tab:servers}
\begin{tabular}{llll}
\toprule
Server (tools) & Provides & Addressing scheme & Example tool call \\
\midrule
\texttt{piop1} (54) & classical atlas time-series, & MNI seed name, orientation & \texttt{piop1\_extract\_roi\_timeseries()} \\
\emph{always on}    & GLM contrasts, graph metrics & only (\texttt{frontal\_inf\_l}) & \\
\addlinespace
\texttt{pubmed} (8) & literature search and & --- (no spatial addressing) & \texttt{pubmed\_search\_articles(query=\ldots)} \\
\emph{always on}    & full-text retrieval & & \\
\addlinespace
\texttt{brainlm} (10) & BrainLM reconstruction error, & named A424 cortical parcel & \texttt{brainlm\_get\_reconstruction\_error(} \\
\emph{ablated}        & forecasting, state similarity & (\texttt{L\_44}) & \quad\texttt{region="L\_44")} \\
\addlinespace
\texttt{omni\_fmri} (11) & Omni-fMRI patch embeddings, & raw voxel-patch coordinate & \texttt{omni\_fmri\_get\_patch\_embedding\_at(} \\
\emph{ablated}           & reconstruction error maps & $(x,y,z)$, atlas-free & \quad\texttt{x=48, y=48, z=48)} \\
\bottomrule
\end{tabular}
\end{table}

\section{Design and robustness tables}
\label{app:tables}
\begin{itemize}
  \item T4: per-criterion differences and which survive correction --
        the numbers behind Fig.~\ref{fig:headline}a.
  \item T6: missing-data sensitivity across the three imputations. The
        primary effect is essentially unchanged under the worst case the
        data admits, so it does not rest on how the 53 unscored runs are
        treated.
\end{itemize}
\begin{table}[t]
\centering
\caption{Equivalence tests (TOST) on paired per-question deltas. Pre-specified margin $\pm0.25$ holistic points (half the CORE primary effect). 90\% CI corresponds to the two one-sided tests at $\alpha=0.05$.}
\label{tab:equivalence}
\begin{tabular}{lrrrrrr}
\toprule
scope & comparison & estimate & ci90 & margin & p\_tost & verdict \\
\midrule
CORE & full vs brainlm only & 0.035 & [-0.11, 0.19] & +/-0.25 & 0.011 & equivalent within +/-0.25 \\
CORE & full vs omni only & 0.253 & [0.06, 0.44] & +/-0.25 & 0.511 & inconclusive (CI spans the margin) \\
all76 & full vs brainlm only & 0.083 & [-0.02, 0.19] & +/-0.25 & 0.005 & equivalent within +/-0.25 \\
all76 & full vs omni only & 0.172 & [0.06, 0.29] & +/-0.25 & 0.132 & inconclusive (CI spans the margin) \\
\bottomrule
\end{tabular}
\end{table}

\begin{table}[H]
\centering
\small
\caption{Where the gain lands. Per-criterion mean difference (both FMs $-$ classical, CORE questions), with Benjamini--Hochberg correction across the ten rubric criteria. The four criteria that describe how the agent \emph{uses} tools all move; the two that describe how it interprets and hedges do not.}
\label{tab:criteria}
\begin{tabular}{llrr}
\toprule
Rubric criterion & Type & $\Delta$ & Survives FDR \\
\midrule
tool selection & tool-facing & +1.09 & yes \\
information sufficiency & tool-facing & +0.85 & yes \\
parameter validity & tool-facing & +0.81 & yes \\
measurement grounding & tool-facing & +0.67 & yes \\
reasoning coherence & interpretive & +0.57 & yes \\
step level reasoning & interpretive & +0.56 & yes \\
neural correlate consistency & interpretive & +0.52 & yes \\
primitive fidelity & interpretive & +0.24 & \textbf{no} \\
epistemic calibration & interpretive & +0.22 & \textbf{no} \\
\bottomrule
\end{tabular}
\end{table}

\begin{table}[H]
\centering
\caption{Missing-data sensitivity for the primary contrast (full $-$ classical). complete-case = surviving runs only; worst-case = $\mathrm{holistic}=1$ for all 53 unscored runs; question-floor = each unscored run imputed at that question's minimum observed score across conditions.}
\label{tab:missing_data}
\begin{tabular}{llrrrrrrr}
\toprule
 & & & & \multicolumn{2}{c}{95\% CI} & & & \\
\cmidrule(lr){5-6}
imputation & scope & $n$ & $\overline{\Delta}$ & lower & upper & $p_{\mathrm{perm}}$ & $p_{\mathrm{Wilcoxon}}$ & $d_{z}$ \\
\midrule
complete case & CORE & 34 & 0.695 & 0.444 & 0.973 & $<0.001$ & $<0.001$ & 0.861 \\
complete case & all-76 & 76 & 0.516 & 0.315 & 0.715 & $<0.001$ & $<0.001$ & 0.575 \\
worst case & CORE & 34 & 0.673 & 0.402 & 0.957 & $<0.001$ & $<0.001$ & 0.816 \\
worst case & all-76 & 76 & 0.481 & 0.271 & 0.682 & $<0.001$ & $<0.001$ & 0.522 \\
question floor & CORE & 34 & 0.673 & 0.410 & 0.951 & $<0.001$ & $<0.001$ & 0.816 \\
question floor & all-76 & 76 & 0.481 & 0.271 & 0.686 & $<0.001$ & $<0.001$ & 0.522 \\
\bottomrule
\end{tabular}
\end{table}

\section{Question-directedness audit}
\label{app:directedness}
\begin{itemize}
  \item Method: tag each question by which foundation model its \emph{text}
        names (case-insensitive \texttt{brainlm} / \texttt{omni} match), then
        re-run every model-comparison contrast within tag. Script:
        \texttt{ANALYSIS\_V2\_directedness.py}; output:
        \texttt{directedness\_v2.json}, \texttt{directedness.csv}.
  \item Counts, all 76 questions: 12 BrainLM-directed, 1 Omni-directed,
        33 symmetric, 30 neutral. CORE: 10 / 1 / 20 / 3.
  \item FM-access headline by tag (full $-$ classical): BrainLM-directed
        $+0.939$ ($p{=}0.007$), symmetric $+0.603$ ($p{=}0.004$), neutral
        $+0.500$ ($n{=}3$) -- C1 holds within every tag.
  \item Ungradeable runs by tag: of \texttt{omni\_only}'s 33, only 2 are on
        BrainLM-directed questions (20 neutral, 11 symmetric), so the
        completion-rate result is unconfounded.
  \item Equivalence by scope: full $-$ brainlm\_only is equivalent on CORE
        ($p_{\mathrm{TOST}}{=}0.011$) and on CORE-symmetric (0.037), but on
        the all-76 symmetric set both contrasts are inconclusive with
        near-equal estimates ($+0.145$ vs.\ $+0.167$).
\end{itemize}

\begin{table}[H]
\centering
\small
\caption{The head-to-head gap is an artifact of question wording. Questions are grouped by which model their text names. The gap lives entirely in questions that name BrainLM's tools, which an \texttt{omni\_only} agent cannot call; where both models are named equally, no difference is detectable.}
\label{tab:directedness}
\begin{tabular}{lrrrr}
\toprule
Question subset & $n$ & BrainLM $-$ Omni & 95\% CI & $p$ \\
\midrule
Names BrainLM's tools & 10 & +0.68 & [+0.34, +1.29] & 0.007 \\
Names both equally & 20 & +0.15 & [-0.17, +0.51] & 0.398 \\
Names Omni's tools & 1 & -1.40 & --- & --- \\
All CORE questions & 34 & +0.22 & [-0.03, +0.52] & 0.136 \\
\bottomrule
\end{tabular}
\end{table}

\section{Reproducibility}
\label{app:stats}
Every number in this paper is emitted by a single analysis notebook into
\texttt{results\_v2.json}: all contrasts on both scopes, BCa bootstrap and
permutation settings (10{,}000 each, seed 0), model fits (ordinal full
OR 2.167, 95\% CI [1.795, 2.618]; proportional-odds spread 0.97), TOST
detail, per-criterion FDR, and the missing-data grid. Of the 3{,}040 runs,
53 produced no gradeable answer (15 timeout, 38 error), distributed
\{classical 3, \texttt{brainlm\_only} 4, \texttt{omni\_only} 33,
\texttt{full} 13\}.

\section{Example traces: one question across the four conditions}
\label{app:traces}
This appendix shows what the ablation actually changes at the level of a single
run. We take VC-07, a Vocabulary-Crossing question chosen because it makes the
addressing problem explicit rather than incidental --- it instructs the agent to
call BrainLM with a region name taken from PIOP1's vocabulary, report the error,
then recover via BrainLM's own discovery call:

\begin{quote}\small
``Deliberately call \texttt{brainlm\_get\_reconstruction\_error} with a region
name copied verbatim from \texttt{piop1\_describe\_dataset()}'s output rather
than from \texttt{brainlm\_list\_brainlm\_regions()}. Report the exact error you
get, if any. Then correct course: call \texttt{brainlm\_list\_brainlm\_regions()}
and retry with a valid name. Report both outcomes and state what the discrepancy
tells you about why per-server discovery calls matter.''
\end{quote}

\noindent
All four traces below are the same question at the same seed (subject~8), so the
only difference between them is which servers were mounted. They are generated
directly from the run logs by \texttt{make\_traces\_appendix.py}, and interleave
the agent's own text with its tool calls in the order they occurred: a
\texttt{[model]} block is what the agent wrote between calls, and the closing
block is the answer it returned. The calls themselves come from
\texttt{tool\_call} records, which carry the harness's own request and response,
so a call the agent merely \emph{claims} in its text cannot appear as a call
here --- a distinction that matters below. Tool \emph{responses} are abridged to
one line (the rejected calls echo the whole 60-odd-tool manifest, and the region
listing is 424 names); the agent's own text is verbatim and complete. This seed's holistic scores (2 / 4 / 3 / 5) track the
per-condition means for VC-07 (2.1 / 4.0 / 3.4 / 4.7, Table~\ref{tab:all_questions}),
but a single run is an illustration of the mechanism, not evidence for the
effect size.

\subsection{\texttt{classical} --- the two foundation-model servers withheld}
\noindent Servers mounted: \texttt{piop1}, \texttt{pubmed}. Judged \texttt{holistic\_score} 2; 4 tool calls in 17.0\,s.
\nopagebreak
\begingroup\footnotesize
\begin{verbatim}
  piop1_describe_dataset()
    -> Regions (MNI seed names, orientation only): ['frontal_inf_l', 'temporal_sup_l',
       'occipital_sup_l', 'putamen_l']
  pubmed_search_articles(query='fMRI brain regions', max_results=5)
    -> 5 results
  brainlm_get_reconstruction_error(region='Left-Cerebellum-Crus1')
    -> Error: brainlm_get_reconstruction_error is not a valid tool, try one of [62
       other tools]

  [model, between tool calls]
    The error message indicates that `brainlm_get_reconstruction_error` is not a valid
      tool. This discrepancy arises because the region names listed in
      `piop1_describe_dataset()` are for orientation purposes and do not correspond to
      the regions recognized by the `brainlm` tools.

    To correct this, we need to use the correct region names from
      `brainlm_list_brainlm_regions()`. Let's proceed with the correct approach:

    1. **List the valid BrainLM regions**:
       ```json
       {
         "name": "brainlm_list_brainlm_regions",
         "arguments": {}
       }
       ```

    2. **Get the reconstruction error for a valid region**:
       ```json
       {
         "name": "brainlm_get_reconstruction_error",
         "arguments": {
           "region": "valid_region_name"  // Replace with an actual valid region name
      from the previous step
         }
       }
       ```

    ### Explanation:
    The discrepancy highlights the importance of using the correct discovery calls
      specific to each server or toolset. The region names provided by
      `piop1_describe_dataset()` are for general orientation and do not match the
      specific regions recognized by the `brainlm` tools. Therefore, it is crucial to
      use `brainlm_list_brainlm_regions()` to get the valid region names for
      `brainlm_get_reconstruction_error`.

    Let's proceed with the correct steps.

  brainlm_list_brainlm_regions()
    -> Error: brainlm_list_brainlm_regions is not a valid tool, try one of [62 other
       tools]

  [final answer]
    The errors encountered highlight a critical point about the importance of using
      the correct tools and functions specific to the server or platform you are
      working with. Here’s a detailed breakdown of the outcomes and their
      implications:

    ### Outcomes:

    1. **Attempt to Call `brainlm_get_reconstruction_error`:**
       - **Error Message:** `brainlm_get_reconstruction_error is not a valid tool, try
      one of [list of valid tools].`
       - **Region Name Used:** `Left-Cerebellum-Crus1`
       - **Source of Region Name:** Copied verbatim from `piop1_describe_dataset()`'s
      output.

    2. **Attempt to Call `brainlm_list_brainlm_regions`:**
       - **Error Message:** `brainlm_list_brainlm_regions is not a valid tool, try one
      of [list of valid tools].`

    ### Implications:

    - **Tool Availability:** The errors indicate that the tools
      `brainlm_get_reconstruction_error` and `brainlm_list_brainlm_regions` are not
      available on the current server. This suggests that the server's API or toolset
      does not include these specific functions.

    - **Region Naming Discrepancy:** The region name `Left-Cerebellum-Crus1` was taken
      from the output of `piop1_describe_dataset()`, but this does not guarantee that
      it is valid for the `brainlm_get_reconstruction_error` tool. The discrepancy
      arises because `piop1_describe_dataset()` provides region names for orientation
      purposes, not necessarily for direct use with other tools.

    - **Per-Server Discovery Calls:** The errors underscore the importance of using
      discovery calls specific to the server you are interacting with. Each server may
      have a different set of available tools and functions. Using the correct
      discovery calls ensures that you are aware of the valid tools and region names
      that can be used.

    ### Conclusion:

    To proceed correctly, you should:

    1. **Identify Valid Tools:** Use the server's discovery calls to list available
      tools and functions. For example, you can use `piop1_list_handles` to see what
      handles are available and `piop1_describe_dataset` to understand the dataset
      structure.

    2. **Use Correct Region Names:** Ensure that the region names you use are valid
      for the specific tool you are calling. If `brainlm_get_reconstruction_error` is
      not available, you may need to use an alternative tool or approach to achieve
      your goal.

    3. **Consult Documentation:** Refer to the server's documentation or help
      resources to understand the correct usage of tools and functions.

    By following these steps, you can avoid errors and ensure that you are using the
      correct tools and region names for your analysis.

  [4 tool calls, 17.0 s, holistic_score = 2]
\end{verbatim}
\endgroup

\subsection{\texttt{brainlm\_only} --- atlas-indexed foundation model only}
\noindent Servers mounted: \texttt{piop1}, \texttt{brainlm}, \texttt{pubmed}. Judged \texttt{holistic\_score} 4; 5 tool calls in 9.1\,s.
\nopagebreak
\begingroup\footnotesize
\begin{verbatim}
  piop1_describe_dataset()
    -> Regions (MNI seed names, orientation only): ['frontal_inf_l', 'temporal_sup_l',
       'occipital_sup_l', 'putamen_l']
  brainlm_get_brainlm_info()
    -> {model: BrainLM (ViT-MAE masked autoencoder), atlas: A424 (424 regions)}
  brainlm_list_brainlm_regions()
    -> 424 names: R_V1, R_MST, R_V6, R_V2, R_V3, R_V4, ...
  brainlm_get_reconstruction_error(start_sec=0, region='frontal_inf_l')
    -> Error executing tool get_reconstruction_error: Unknown region 'frontal_inf_l'.
       Call list_brainlm_regions() for valid names.
  brainlm_get_reconstruction_error(start_sec=0, region='L_V1')
    -> {value: 0.16537953913211823}

  [final answer]
    The first call to `brainlm_get_reconstruction_error` with the region name
      `frontal_inf_l` (copied from `piop1_describe_dataset()`) resulted in an error:

    ```
    Error executing tool get_reconstruction_error: Unknown region 'frontal_inf_l'.
      Call list_brainlm_regions() for valid names.
    ```

    This error occurred because `frontal_inf_l` is not a valid region name for the
      BrainLM tool. The valid region names are listed by
      `brainlm_list_brainlm_regions()`, which includes names like `L_V1`.

    After correcting the region name to `L_V1` (a valid region name from
      `brainlm_list_brainlm_regions()`), the call succeeded and returned the
      reconstruction error value:

    ```
    {
      "value": 0.16537953913211823,
      "note": null
    }
    ```

    This discrepancy highlights the importance of using the correct discovery calls
      specific to each server or tool. The region names listed in
      `piop1_describe_dataset()` are for orientation and metadata purposes, while the
      BrainLM tool requires region names from `brainlm_list_brainlm_regions()`. Using
      the wrong set of region names can lead to errors, as demonstrated.

  [5 tool calls, 9.1 s, holistic_score = 4]
\end{verbatim}
\endgroup

\subsection{\texttt{omni\_only} --- voxel-indexed foundation model only}
\noindent Servers mounted: \texttt{piop1}, \texttt{omni\_fmri}, \texttt{pubmed}. Judged \texttt{holistic\_score} 3; 3 tool calls in 17.3\,s.
\nopagebreak
\begingroup\footnotesize
\begin{verbatim}
  piop1_describe_dataset()
    -> Regions (MNI seed names, orientation only): ['frontal_inf_l', 'temporal_sup_l',
       'occipital_sup_l', 'putamen_l']
  omni_fmri_get_omnifmri_info()
    -> {model: Omni-fMRI (atlas-free AdaptiveMAE, dynamic 3D patching), atlas: null,
       note_atlas_free: No fixed region parcellation -- analysis addresses voxel-s...}

  [model, between tool calls]
    The `piop1_describe_dataset` function returns the following regions:
      `['frontal_inf_l', 'temporal_sup_l', 'occipital_sup_l', 'putamen_l']`.

    However, these regions are not valid for the `brainlm_get_reconstruction_error`
      function. To find valid regions, we need to call `brainlm_list_brainlm_regions`.

    Let's proceed with calling `brainlm_list_brainlm_regions` to get the valid region
      names.

  brainlm_list_brainlm_regions()
    -> Error: brainlm_list_brainlm_regions is not a valid tool, try one of [73 other
       tools]

  [final answer]
    The error message indicates that `brainlm_list_brainlm_regions` is not a valid
      tool. This suggests that the tool name was incorrectly referenced. Instead, we
      should use the correct tool names available in the list provided.

    Given the context, it seems there was a misunderstanding about the available
      tools. The correct approach is to use the tools provided by the `piop1` and
      `omni_fmri` modules.

    Let's correct the course by using the correct tool names. Since
      `brainlm_list_brainlm_regions` is not a valid tool, we need to identify the
      correct tool to list the regions. However, based on the available tools, it
      seems there is no direct equivalent to `brainlm_list_brainlm_regions` in the
      `piop1` or `omni_fmri` modules.

    Instead, we can use `piop1_describe_dataset` to get the regions, but this will not
      provide the same detailed region information as `brainlm_list_brainlm_regions`.
      The `omni_fmri` module does not provide region information as it is atlas-free.

    ### Outcome of the Incorrect Call
    Attempting to call `brainlm_get_reconstruction_error` with a region name from
      `piop1_describe_dataset` would result in an error because the region names
      provided by `piop1_describe_dataset` are not compatible with the `brainlm`
      tools. The error message would indicate that the region name is invalid.

    ### Correct Course of Action
    Since `brainlm_list_brainlm_regions` is not a valid tool, we cannot list the
      regions directly. However, we can use the `piop1_describe_dataset` tool to get
      an orientation of the dataset, including the regions. Here is the corrected
      approach:

    1. Use `piop1_describe_dataset` to get the regions.
    2. Use the correct tool names from the provided list.

    ### What the Discrepancy Tells Us
    The discrepancy highlights the importance of using the correct tool names and
      understanding the capabilities of each server. The `piop1` and `omni_fmri`
      modules have different functionalities and region handling mechanisms. The
      `piop1` module provides region information based on the AAL-424 atlas, while the
      `omni_fmri` module is atlas-free and focuses on voxel-space patch coordinates.
      This underscores the need for per-server discovery calls to ensure that the
      correct tools and region names are used.

  [3 tool calls, 17.3 s, holistic_score = 3]
\end{verbatim}
\endgroup

\subsection{\texttt{full} --- both foundation models}
\noindent Servers mounted: \texttt{piop1}, \texttt{brainlm}, \texttt{omni\_fmri}, \texttt{pubmed}. Judged \texttt{holistic\_score} 5; 6 tool calls in 21.4\,s.
\nopagebreak
\begingroup\footnotesize
\begin{verbatim}
  piop1_describe_dataset()
    -> Regions (MNI seed names, orientation only): ['frontal_inf_l', 'temporal_sup_l',
       'occipital_sup_l', 'putamen_l']
  brainlm_get_brainlm_info()
    -> {model: BrainLM (ViT-MAE masked autoencoder), atlas: A424 (424 regions)}
  brainlm_list_brainlm_regions()
    -> 424 names: R_V1, R_MST, R_V6, R_V2, R_V3, R_V4, ...
  omni_fmri_get_omnifmri_info()
    -> {model: Omni-fMRI (atlas-free AdaptiveMAE, dynamic 3D patching), atlas: null,
       note_atlas_free: No fixed region parcellation -- analysis addresses voxel-s...}

  [model, between tool calls]
    To demonstrate the importance of using the correct region names from the
      appropriate discovery call, I will first call `brainlm_get_reconstruction_error`
      with a region name copied verbatim from `piop1_describe_dataset()`'s output.
      Then, I will correct the region name using `brainlm_list_brainlm_regions()` and
      retry the call.

    ### Step 1: Call `brainlm_get_reconstruction_error` with an incorrect region name

    Let's use the region name `frontal_inf_l` from `piop1_describe_dataset()`'s
      output.

    ```json
    {
      "name": "brainlm_get_reconstruction_error",
      "arguments": {
        "start_sec": 0.0,
        "region": "frontal_inf_l"
      }
    }
    ```

    ### Step 2: Correct the region name and retry

    Now, let's use a valid region name from `brainlm_list_brainlm_regions()`'s output.
      For example, we can use `L_V1`.

    ```json
    {
      "name": "brainlm_get_reconstruction_error",
      "arguments": {
        "start_sec": 0.0,
        "region": "L_V1"
      }
    }
    ```

    ### Expected Outcomes

    1. **Incorrect Region Name**:
       - **Error**: The call with the incorrect region name `frontal_inf_l` will
      likely fail with an error indicating that the region name is not recognized.

    2. **Correct Region Name**:
       - **Success**: The call with the correct region name `L_V1` will succeed and
      return the reconstruction error for that region.

    ### Explanation

    The discrepancy between the region names returned by `piop1_describe_dataset()`
      and `brainlm_list_brainlm_regions()` highlights the importance of using the
      correct discovery calls for each server. The `piop1_describe_dataset()` function
      provides general information about the dataset, including region names that are
      useful for orientation but may not match the specific region names required by
      the BrainLM server. The `brainlm_list_brainlm_regions()` function provides the
      exact region names that the BrainLM server expects, ensuring that calls to
      BrainLM functions are successful.

    Let's proceed with the actual calls to observe the outcomes.

  brainlm_get_reconstruction_error(start_sec=0, region='frontal_inf_l')
    -> Error executing tool get_reconstruction_error: Unknown region 'frontal_inf_l'.
       Call list_brainlm_regions() for valid names.
  brainlm_get_reconstruction_error(start_sec=0, region='L_V1')
    -> {value: 0.14288929104804993}

  [final answer]
    The first call to `brainlm_get_reconstruction_error` with the region name
      `"frontal_inf_l"` (copied from `piop1_describe_dataset()`) resulted in an error:

    ```
    Error executing tool get_reconstruction_error: Unknown region 'frontal_inf_l'.
      Call list_brainlm_regions() for valid names.
    ```

    The second call to `brainlm_get_reconstruction_error` with the region name
      `"L_V1"` (copied from `brainlm_list_brainlm_regions()`) succeeded and returned
      the reconstruction error value:

    ```
    {
      "value": 0.14288929104804993,
      "note": null
    }
    ```

    This discrepancy highlights the importance of using the correct region names
      specific to the BrainLM tool. The region names returned by
      `piop1_describe_dataset()` are not the same as those expected by the BrainLM
      tools. Therefore, it is crucial to call `brainlm_list_brainlm_regions()` to get
      the valid region names for BrainLM-specific operations. This ensures that the
      region names used are compatible with the BrainLM tool, avoiding errors and
      ensuring accurate results.

  [6 tool calls, 21.4 s, holistic_score = 5]
\end{verbatim}
\endgroup

\paragraph{What the four traces show.}
Only the two conditions with BrainLM mounted can execute the question at all.
In \texttt{classical} and \texttt{omni\_only} the BrainLM tools are not merely
unhelpful, they are absent: the harness rejects the call by name before any
region string is ever parsed, so the vocabulary error the question is about
cannot be reached. In \texttt{brainlm\_only} and \texttt{full} the same call is
dispatched and fails for the intended reason --- \texttt{frontal\_inf\_l} is a
PIOP1 MNI seed name and is not among BrainLM's 424 A424 parcels --- and the
agent recovers through \texttt{brainlm\_list\_brainlm\_regions()} to a valid
name, \texttt{L\_V1}, returning a real measurement. This is the substitution the
paper's addressing argument rests on, visible in one run.

\paragraph{Two failures the score does not catch.}
The judged score rewards a coherent account of what happened, and both
FM-less conditions supply one without any grounding. The
\texttt{classical} answer reports that it used the region name
\texttt{Left-Cerebellum-Crus1} ``copied verbatim from
\texttt{piop1\_describe\_dataset()}'s output'' --- that output listed four names,
and this was not one of them. The \texttt{omni\_only} answer describes the error
it \emph{would} have received (``Attempting to call
\texttt{brainlm\_get\_reconstruction\_error} \ldots would result in an error'')
and asserts that PIOP1 ``provides region information based on the AAL-424
atlas'', conflating PIOP1's orientation-only seed names with BrainLM's parcel
atlas. Neither run was penalised much for this: they scored 2 and 3 on a scale
where 1 is reserved for incoherence. This is the same asymmetry
Table~\ref{tab:criteria} reports in aggregate --- the criteria describing how an
agent \emph{uses} tools move under FM access, while
\texttt{primitive\_fidelity} and \texttt{epistemic\_calibration}, which describe
how it reads a value back, do not.

\clearpage
\section{All 76 questions}
\label{app:questions}
\begingroup\scriptsize\setlength{\tabcolsep}{2.5pt}
\begin{longtable}{@{}llp{1.9cm}rrrrr@{}}
\caption{All 76 benchmark questions: qid, family, task stage, per-condition mean holistic score, number of scored runs (out of 40), and the full prompt text as issued to the agent.}\label{tab:all_questions}\\
\toprule
qid & family & stage & classical & \texttt{brainlm} & \texttt{omni} & full & $n$ \\
\midrule
\endfirsthead
\toprule
qid & family & stage & classical & \texttt{brainlm} & \texttt{omni} & full & $n$ \\
\midrule
\endhead
\bottomrule
\endfoot
\texttt{CB-01} & CB & anticipation & 4.444 & 4.222 & 3.625 & 4.111 & 35 \\*[1pt]
\multicolumn{8}{@{}p{13.4cm}@{}}{\raggedright\footnotesize PIOP1's own discovery tool names a region 'frontal\_\allowbreak{}inf\_\allowbreak{}l' -- note this is an MNI seed name for orientation only, not a formal atlas parcel. Call BrainLM's region-listing tool and give your top three candidate BrainLM region names for that same rough territory, ranked by confidence. For each candidate, state one piece of evidence a tool could return that would confirm it and one that would rule it out. Do not present a single name as definitively correct -- the correspondence is inherently approximate, and your answer should reflect that.} \\[2pt]
\qsep
\texttt{CB-02} & CB & workingmemory & 3.889 & 4.200 & 4.000 & 4.200 & 39 \\*[1pt]
\multicolumn{8}{@{}p{13.4cm}@{}}{\raggedright\footnotesize Working-memory delay-period analysis needs a comparable region between BrainLM's named-region tool and Omni-fMRI's voxel-coordinate tool, but the two servers share no vocabulary and PIOP1's own region names are seed locations, not atlas parcels. Give your top two BrainLM region candidates and top two Omni-fMRI voxel-coordinate candidates for the same rough prefrontal territory, explicitly rank them by how confident you are in the correspondence, and state what would change your ranking (e.g. a specific reconstruction-error pattern, a specific coordinate range) rather than asserting a single best-guess pair as fact.} \\[2pt]
\qsep
\texttt{CB-03} & CB & emomatching & 4.375 & 4.600 & 4.400 & 4.667 & 37 \\*[1pt]
\multicolumn{8}{@{}p{13.4cm}@{}}{\raggedright\footnotesize Identify the emotion-processing-relevant region from PIOP1's own region/condition metadata, then give your top two BrainLM region candidates for the same anatomical territory rather than a single pick. For each candidate, state explicitly whether you are confident enough in the name correspondence to treat a BrainLM reconstruction-error result at that region as evidence about the PIOP1-named region specifically, or whether the mapping is too approximate to support that inference.} \\[2pt]
\qsep
\texttt{EB-01} & EB & gstroop & 4.700 & 4.400 & 4.600 & 4.400 & 40 \\*[1pt]
\multicolumn{8}{@{}p{13.4cm}@{}}{\raggedright\footnotesize Search PubMed for the exact phrase 'reconstruction-error breakdown point under masking-fraction sweep in fMRI foundation models'. You have a budget of 3 tool calls total for this question -- plan your search terms before calling, and if the first search returns nothing, do not keep retrying minor rewordings; report the null result and stop. This is a narrow, purpose-built phrase unlikely to match a real paper title.} \\[2pt]
\qsep
\texttt{EB-02} & EB & restingstate & 2.875 & 3.000 & 3.500 & 4.100 & 36 \\*[1pt]
\multicolumn{8}{@{}p{13.4cm}@{}}{\raggedright\footnotesize Sweep the masking fraction from light to aggressive in both foundation models to see whether their reconstruction cost rises at a similar rate. You have a budget of 10 tool calls total across both models -- decide up front how many mask ratios per model that budget allows, state your plan before calling anything, then execute it. End your response with exactly one line: VERDICT: $<$SIMILAR$|$DIVERGENT$>$ -- BrainLM slope/trend: $<$description$>$; Omni-fMRI slope/trend: $<$description$>$.} \\[2pt]
\qsep
\texttt{EB-03} & EB & any\_\allowbreak{}task & 3.700 & 3.600 & 4.100 & 3.900 & 40 \\*[1pt]
\multicolumn{8}{@{}p{13.4cm}@{}}{\raggedright\footnotesize Do this in four phases, and you have a total budget of 15 tool calls across all four phases combined -- state how you're allocating that budget across the phases before you start. Phase one: compute this run's classical connectivity hubs. Phase two: sample BrainLM reconstruction error across the run to find its top anomalous window. Phase three: do the same in Omni-fMRI. Phase four: state whether you stayed within budget, and if you didn't, say which phase overran and why.} \\[2pt]
\qsep
\texttt{EN-01} & EN & workingmemory & 4.200 & 4.300 & 4.500 & 4.500 & 38 \\*[1pt]
\multicolumn{8}{@{}p{13.4cm}@{}}{\raggedright\footnotesize Do this in five phases. Phase one: retrieve every active-change and active-nochange trial onset and delay-period window. Phase two: run the full classical graph-theoretic pipeline (threshold, betweenness centrality, modularity) on this run's connectivity and report the top three hub regions -- this phase alone is a complete, publishable classical analysis. Phase three: at those same three hub regions, get BrainLM reconstruction error during the delay period for both trial types. Phase four: get Omni-fMRI reconstruction error at the corresponding voxel coordinates for the same trials. Phase five: state specifically what phases three and four added beyond what phase two's classical analysis alone already told you -- do not simply restate that the FM results exist; say whether they sharpened, contradicted, or were redundant with the classical hub finding.} \\[2pt]
\qsep
\texttt{EN-02} & EN & gstroop & 4.333 & 4.100 & 4.600 & 4.222 & 38 \\*[1pt]
\multicolumn{8}{@{}p{13.4cm}@{}}{\raggedright\footnotesize Do this in five phases. Phase one: compute the classical incongruent-vs-congruent contrast and report its peak region and magnitude -- a complete classical answer to 'where does conflict monitoring show up.' Phase two: at that exact peak region, get BrainLM reconstruction error separately for congruent and incongruent trials. Phase three: at the corresponding voxels, get Omni-fMRI reconstruction error the same way. Phase four: run a mask-ratio sweep on incongruent trials in both models at the peak region/voxels specifically (not the whole brain). Phase five: state whether the FM evidence in phases two through four sharpens the classical peak from phase one (e.g. confirms it's harder to predict, not just more active), or whether it's just a redundant restatement of what the contrast already showed.} \\[2pt]
\qsep
\texttt{EN-03} & EN & emomatching & 3.667 & 4.500 & 4.400 & 4.200 & 39 \\*[1pt]
\multicolumn{8}{@{}p{13.4cm}@{}}{\raggedright\footnotesize Do this in five phases. Phase one: compute the classical emotion-vs-control contrast and report its peak region -- a complete classical answer on its own. Phase two: at that peak region, get BrainLM reconstruction error split by condition (emotion vs control), reporting each trial. Phase three: at the corresponding voxels, get Omni-fMRI reconstruction error the same way. Phase four: search the literature for whether emotional content is expected to be more or less predictable to a learned model, reporting exactly what the search returns. Phase five: state whether phases two through four sharpen, contradict, or are redundant with the classical contrast from phase one -- and separately, whether the literature from phase four actually supports whatever direction your own data showed, not what you assumed it would say.} \\[2pt]
\qsep
\texttt{ES-01} & ES & anticipation & 5.000 & 4.889 & 4.100 & 4.400 & 39 \\*[1pt]
\multicolumn{8}{@{}p{13.4cm}@{}}{\raggedright\footnotesize Do this in five phases, and after each phase state whether a classical GLM/contrast tool could have produced that phase's result on its own. Phase one: retrieve every cue/image trial onset for this run. Phase two: for the cue period of each trial, get BrainLM's forecast-difficulty value predicting forward into the image period, reporting every trial individually -- note explicitly that this forward-prediction measurement has no classical GLM equivalent. Phase three: get Omni-fMRI's own forecast-difficulty value for the same cue-to-image transitions, again per trial. Phase four: compute the classical cue-vs-image contrast and its peak region, for comparison only. Phase five: state plainly which of phases two through four could only have been produced by a foundation model, and which could have been produced by classical tools alone.} \\[2pt]
\qsep
\texttt{ES-02} & ES & restingstate & 2.400 & 3.500 & 4.100 & 4.111 & 39 \\*[1pt]
\multicolumn{8}{@{}p{13.4cm}@{}}{\raggedright\footnotesize Do this in five phases. Phase one: compute this run's classical connectivity matrix and identify the top three hub regions by betweenness centrality -- this is a complete classical answer to 'what regions matter most' on its own. Phase two: separately, without using the connectivity matrix at all, sample at least eight windows across the run in BrainLM and report each window's raw reconstruction error, to find the model's own top candidate for the most atypical window. Phase three: do the same for Omni-fMRI. Phase four: state explicitly whether 'most atypical window in time' (phases two-three) is even the same kind of finding as 'most central hub region in space' (phase one), or whether classical connectivity analysis structurally cannot answer the temporal-anomaly question at all. Phase five: give your final position on whether phases two and three revealed something phase one could never have found, regardless of whether the specific regions/windows agree.} \\[2pt]
\qsep
\texttt{ES-03} & ES & any\_\allowbreak{}task & 3.556 & 4.111 & 4.250 & 4.375 & 34 \\*[1pt]
\multicolumn{8}{@{}p{13.4cm}@{}}{\raggedright\footnotesize Do this in five phases, treating this run generically since it has no task-specific event structure. Phase one: state directly what a classical GLM/contrast analysis needs to run (a design matrix, task events, conditions to contrast) and confirm none of that exists for this run. Phase two: get BrainLM's embedding-space state-similarity between at least four evenly-spaced windows across the run, reporting every pairwise value. Phase three: do the same in Omni-fMRI. Phase four: identify, from phases two and three alone, which two windows this subject's brain state changed most between. Phase five: state explicitly that this specific finding (a state-space transition point with no external task label to explain it) is a capability classical GLM tools cannot produce at all, and say why the absence of task events is exactly what makes the FM-only approach necessary here rather than optional.} \\[2pt]
\qsep
\texttt{LC-01} & LC & any\_\allowbreak{}task & 3.400 & 3.400 & 3.400 & 3.600 & 40 \\*[1pt]
\multicolumn{8}{@{}p{13.4cm}@{}}{\raggedright\footnotesize Search PubMed for a direct empirical comparison between 'BrainLM' and 'Omni-fMRI' as named models. Report exactly what the search tool returns. If it returns nothing, or nothing that actually names both models, say so plainly -- do not construct a title, author list, or PMID that the tool did not give you.} \\[2pt]
\qsep
\texttt{LC-02} & LC & anticipation & 3.100 & 2.700 & 1.700 & 2.900 & 40 \\*[1pt]
\multicolumn{8}{@{}p{13.4cm}@{}}{\raggedright\footnotesize Every PMID or paper title in your final answer must be one that a pubmed tool call in this conversation actually returned verbatim. Search the literature on anticipatory cue-locked prediction error, and report only what the search actually returned, including if it returned zero results.} \\[2pt]
\qsep
\texttt{LC-03} & LC & gstroop & 4.500 & 3.857 & 3.875 & 2.833 & 31 \\*[1pt]
\multicolumn{8}{@{}p{13.4cm}@{}}{\raggedright\footnotesize Search PubMed for the exact phrase 'reconstruction-error breakdown point under masking-fraction sweep in fMRI foundation models'. This is a narrow, purpose-built phrase unlikely to match a real paper title. Report the raw search result count and whatever the tool returned, and explicitly state whether any hit is a genuine match to that phrase or only a loose keyword overlap.} \\[2pt]
\qsep
\texttt{LC-04} & LC & gstroop & 3.700 & 3.400 & 3.556 & 4.222 & 38 \\*[1pt]
\multicolumn{8}{@{}p{13.4cm}@{}}{\raggedright\footnotesize Run a PubMed search on the Stroop task and anterior cingulate cortex conflict monitoring -- a well-established literature that should return real hits. Report the actual titles and PMIDs the tool returns, then separately run a second search combining this run's specific model checkpoint filename with 'comparison study', which is unlikely to return anything real. Report both result sets side by side and flag which search came back empty or near-empty.} \\[2pt]
\qsep
\texttt{LC-05} & LC & emomatching & 3.500 & 3.100 & 3.900 & 3.800 & 40 \\*[1pt]
\multicolumn{8}{@{}p{13.4cm}@{}}{\raggedright\footnotesize Search for published work on emotional content and neural predictability relevant to this run's emotion-matching design. Before writing your final answer, list every PMID you intend to cite and confirm each one appears in a tool result earlier in this conversation. If a claim you want to make has no supporting search result, mark it as your own inference rather than attributing it to the literature.} \\[2pt]
\qsep
\texttt{LC-06} & LC & faces & 4.200 & 3.700 & 4.000 & 3.300 & 40 \\*[1pt]
\multicolumn{8}{@{}p{13.4cm}@{}}{\raggedright\footnotesize Search the literature for expression-specific neural reliability findings relevant to this run's facial-expression blocks. Then attempt a second, deliberately overly specific search combining this exact subject's demographic profile with the facial-expression task name. Compare how the specificity of the second query affected whether the tool returned anything at all.} \\[2pt]
\qsep
\texttt{LC-07} & LC & restingstate & 4.100 & 4.600 & 3.778 & 3.667 & 38 \\*[1pt]
\multicolumn{8}{@{}p{13.4cm}@{}}{\raggedright\footnotesize Search PubMed for literature on hub regions and predictability in resting-state functional connectivity. Report the titles and PMIDs actually returned. Then, separately, ask whether the literature you found specifically discusses foundation-model reconstruction error at hubs (as opposed to classical graph metrics) -- if it does not, say that this specific angle is not covered by what you found, rather than extrapolating a citation to cover it.} \\[2pt]
\qsep
\texttt{LC-08} & LC & workingmemory & 3.500 & 3.400 & 3.111 & 3.400 & 39 \\*[1pt]
\multicolumn{8}{@{}p{13.4cm}@{}}{\raggedright\footnotesize Search for literature on working-memory load and neural predictability. If the search tool returns results, pick the single most relevant one and quote its actual title and PMID as returned by the tool, without paraphrasing the title into something that sounds more directly on-topic than what was actually returned.} \\[2pt]
\qsep
\texttt{LC-09} & LC & any\_\allowbreak{}task & 3.625 & 3.556 & 3.600 & 3.700 & 37 \\*[1pt]
\multicolumn{8}{@{}p{13.4cm}@{}}{\raggedright\footnotesize Two searches: first, a broad, well-established topic (the AOMIC PIOP1 dataset itself). Second, a specific but almost certainly nonexistent claim: 'AOMIC PIOP1 anticipation task validated for masked-autoencoder foundation model benchmarking'. Report both raw result sets. For the second, if nothing matches, say directly that this specific validation claim appears to be unsupported by the literature you can find -- do not soften an empty result into a vague-but-confident-sounding citation.} \\[2pt]
\qsep
\texttt{LC-10} & LC & restingstate & 4.000 & 4.500 & 3.400 & 4.400 & 37 \\*[1pt]
\multicolumn{8}{@{}p{13.4cm}@{}}{\raggedright\footnotesize You are asked to defend a specific scientific claim about this run using literature support. Before searching, state the claim. Then search for support. If your search returns strong support, cite it exactly as returned. If it returns weak, tangential, or no support, revise your stated claim down to only what the literature actually backs, and show the before/after revision explicitly.} \\[2pt]
\qsep
\texttt{SX-01A} & SX & faces & 1.000 & 2.222 & 2.100 & 2.100 & 39 \\*[1pt]
\multicolumn{8}{@{}p{13.4cm}@{}}{\raggedright\footnotesize Compare reconstruction difficulty for this run's facial-expression blocks using BrainLM's named-region tool and Omni-fMRI's voxel-coordinate tool. Report both values.} \\[2pt]
\qsep
\texttt{SX-01B} & SX & faces & 2.300 & 3.556 & 3.778 & 3.375 & 36 \\*[1pt]
\multicolumn{8}{@{}p{13.4cm}@{}}{\raggedright\footnotesize Compare reconstruction difficulty for this run's facial-expression blocks using BrainLM's named-region tool and Omni-fMRI's voxel-coordinate tool. Before reporting either value, explicitly state how you chose a BrainLM region name and an Omni-fMRI voxel coordinate that refer to the same physical territory, and confirm neither value leaked from the other tool's vocabulary.} \\[2pt]
\qsep
\texttt{SX-02A} & SX & any\_\allowbreak{}task & 1.700 & 3.000 & 3.444 & 3.300 & 39 \\*[1pt]
\multicolumn{8}{@{}p{13.4cm}@{}}{\raggedright\footnotesize This run has no task-specific design. Identify a region of interest using whichever server's discovery tool you check first, then get reconstruction error for that region from both foundation models.} \\[2pt]
\qsep
\texttt{SX-02B} & SX & any\_\allowbreak{}task & 1.800 & 3.400 & 3.667 & 4.000 & 34 \\*[1pt]
\multicolumn{8}{@{}p{13.4cm}@{}}{\raggedright\footnotesize This run has no task-specific design. Identify a region of interest using whichever server's discovery tool you check first, then get reconstruction error for that region from both foundation models. Before calling either FM tool, state explicitly which server's own vocabulary you are using for each call, and confirm you are not reusing a name or coordinate discovered from one server's tool on a different server's tool.} \\[2pt]
\qsep
\texttt{SX-03A} & SX & restingstate & 2.600 & 4.100 & 3.667 & 4.500 & 37 \\*[1pt]
\multicolumn{8}{@{}p{13.4cm}@{}}{\raggedright\footnotesize Confirm whether BrainLM's atlas and PIOP1's connectivity atlas, both nominally 424 regions, use interchangeable region-name strings, by checking a few names from each.} \\[2pt]
\qsep
\texttt{SX-03B} & SX & restingstate & 2.222 & 4.222 & 4.375 & 4.556 & 35 \\*[1pt]
\multicolumn{8}{@{}p{13.4cm}@{}}{\raggedright\footnotesize Confirm whether BrainLM's atlas and PIOP1's connectivity atlas, both nominally 424 regions, use interchangeable region-name strings, by checking a few names from each. State explicitly, before you conclude either way, exactly which names you compared and what specific difference (or lack of one) you found between them -- do not conclude 'they're the same' or 'they're different' without quoting the actual strings you checked.} \\[2pt]
\qsep
\texttt{SY-01A} & SY & anticipation & 4.000 & 2.100 & 3.111 & 3.111 & 38 \\*[1pt]
\multicolumn{8}{@{}p{13.4cm}@{}}{\raggedright\footnotesize Rank the four cue/image conditions in this run by how surprising each is to a learned model's own prior, separately for each of the two foundation models. Compute both rankings fully before answering. End your response with exactly one line in this form: VERDICT: $<$MATCH$|$MISMATCH$>$ -- BrainLM ranking: $<$order$>$; Omni-fMRI ranking: $<$order$>$.} \\[2pt]
\qsep
\texttt{SY-01B} & SY & anticipation & 4.400 & 4.100 & 3.600 & 3.700 & 40 \\*[1pt]
\multicolumn{8}{@{}p{13.4cm}@{}}{\raggedright\footnotesize Rank the four cue/image conditions in this run by how surprising each is to a learned model's own prior, separately for each of the two foundation models. If the two models' rankings differ, state plainly which model's ranking you would trust more for this specific question and give one concrete reason grounded in how each model represents brain state -- do not simply report both rankings and stop.} \\[2pt]
\qsep
\texttt{SY-02A} & SY & anticipation & 1.700 & 2.000 & 2.000 & 2.300 & 36 \\*[1pt]
\multicolumn{8}{@{}p{13.4cm}@{}}{\raggedright\footnotesize Do the two foundation models agree on which specific cue-to-image transition, valid or invalid, produces the larger jump in learned brain state? Compute the jump for both transition types in both models. End your response with exactly one line: VERDICT: $<$AGREE$|$DISAGREE$>$ -- BrainLM says: $<$transition type$>$; Omni-fMRI says: $<$transition type$>$.} \\[2pt]
\qsep
\texttt{SY-02B} & SY & anticipation & 3.300 & 2.300 & 2.889 & 2.500 & 39 \\*[1pt]
\multicolumn{8}{@{}p{13.4cm}@{}}{\raggedright\footnotesize Do the two foundation models agree on which specific cue-to-image transition, valid or invalid, produces the larger jump in learned brain state? If they disagree, do not just report both answers -- commit to a single best-supported answer for this run and explain in one sentence why you weighted one model's evidence over the other's.} \\[2pt]
\qsep
\texttt{SY-03A} & SY & gstroop & 2.700 & 2.400 & 3.500 & 2.600 & 40 \\*[1pt]
\multicolumn{8}{@{}p{13.4cm}@{}}{\raggedright\footnotesize Do the two foundation models agree on whether incongruent trials are harder to predict in advance than congruent trials? Test this directly in both models. End your response with exactly one line: VERDICT: $<$AGREE$|$DISAGREE$>$ -- BrainLM effect size: $<$value$>$; Omni-fMRI effect size: $<$value$>$.} \\[2pt]
\qsep
\texttt{SY-03B} & SY & gstroop & 2.300 & 3.000 & 3.000 & 3.100 & 36 \\*[1pt]
\multicolumn{8}{@{}p{13.4cm}@{}}{\raggedright\footnotesize Do the two foundation models agree on whether incongruent trials are harder to predict in advance than congruent trials? If one model shows the effect clearly and the other does not, do not leave it as an open question -- state your single best-supported conclusion about whether the underlying effect is real in this subject, and name which model's result you are leaning on and why.} \\[2pt]
\qsep
\texttt{SY-04A} & SY & gstroop & 1.400 & 2.778 & 3.200 & 3.100 & 39 \\*[1pt]
\multicolumn{8}{@{}p{13.4cm}@{}}{\raggedright\footnotesize Both foundation models can independently sweep mask ratio for this run's incongruent trials. Do they agree on whether prediction becomes disproportionately harder past a particular masking level? Run the sweep in both models before answering. End your response with exactly one line: VERDICT: $<$AGREE$|$DISAGREE$>$ -- BrainLM breakdown point: $<$value or none$>$; Omni-fMRI breakdown point: $<$value or none$>$.} \\[2pt]
\qsep
\texttt{SY-04B} & SY & gstroop & 3.400 & 2.600 & 3.000 & 3.300 & 40 \\*[1pt]
\multicolumn{8}{@{}p{13.4cm}@{}}{\raggedright\footnotesize Both foundation models can independently sweep mask ratio for this run's incongruent trials. If only one model shows a clear breakdown point, decide whether that is more likely a real property of the underlying brain data or an artifact of one model's architecture (atlas-based vs voxel-based), and state your reasoning in two sentences, not a list of caveats.} \\[2pt]
\qsep
\texttt{SY-05A} & SY & workingmemory & 1.333 & 3.100 & 2.500 & 2.200 & 39 \\*[1pt]
\multicolumn{8}{@{}p{13.4cm}@{}}{\raggedright\footnotesize Comparing forecast difficulty from one learned model against reconstruction difficulty from the other for active-change trials, do both models point toward the same conclusion about whether change trials are neurally more demanding? End your response with exactly one line: VERDICT: $<$AGREE$|$DISAGREE$>$ -- model A value: $<$value$>$; model B value: $<$value$>$.} \\[2pt]
\qsep
\texttt{SY-05B} & SY & workingmemory & 3.800 & 3.556 & 3.600 & 4.100 & 39 \\*[1pt]
\multicolumn{8}{@{}p{13.4cm}@{}}{\raggedright\footnotesize Comparing forecast difficulty from one learned model against reconstruction difficulty from the other for active-change trials -- two different measurement types, not the same metric -- decide explicitly whether it is even fair to call this 'agreement' given the metrics differ, and then give your best single answer anyway about whether change trials are neurally more demanding in this subject.} \\[2pt]
\qsep
\texttt{SY-06A} & SY & workingmemory & 2.000 & 3.700 & 2.222 & 3.200 & 39 \\*[1pt]
\multicolumn{8}{@{}p{13.4cm}@{}}{\raggedright\footnotesize Considering only the delay period of active trials, do the two foundation models' error profiles -- one region-level, one whole-window -- implicate the same rough part of the brain as most poorly predicted? End your response with exactly one line: VERDICT: $<$SAME$|$DIFFERENT$>$ -- BrainLM region: $<$name$>$; Omni-fMRI location: $<$coordinate$>$.} \\[2pt]
\qsep
\texttt{SY-06B} & SY & workingmemory & 2.600 & 4.200 & 3.000 & 3.500 & 40 \\*[1pt]
\multicolumn{8}{@{}p{13.4cm}@{}}{\raggedright\footnotesize Considering only the delay period of active trials, if the region-level and whole-window error profiles point to different parts of the brain, decide whether that is a genuine localization disagreement or an artifact of one tool averaging over a coarser unit than the other, and say which explanation you believe and why in no more than three sentences.} \\[2pt]
\qsep
\texttt{SY-07A} & SY & emomatching & 2.300 & 1.700 & 2.700 & 2.100 & 40 \\*[1pt]
\multicolumn{8}{@{}p{13.4cm}@{}}{\raggedright\footnotesize Do the two foundation models single out the same trial as the single most atypical one in this emotion-matching run? Score every available trial in both models before answering. End your response with exactly one line: VERDICT: $<$SAME$|$DIFFERENT$>$ -- BrainLM top trial onset: $<$value$>$; Omni-fMRI top trial onset: $<$value$>$.} \\[2pt]
\qsep
\texttt{SY-07B} & SY & emomatching & 2.100 & 2.300 & 3.000 & 2.400 & 39 \\*[1pt]
\multicolumn{8}{@{}p{13.4cm}@{}}{\raggedright\footnotesize Do the two foundation models single out the same trial as the single most atypical one in this emotion-matching run? If their top picks diverge, state which model's pick you find more scientifically defensible for identifying a genuinely atypical brain state, and defend that choice in one sentence tied to the model's architecture.} \\[2pt]
\qsep
\texttt{SY-08A} & SY & faces & 1.900 & 2.300 & 1.857 & 2.200 & 37 \\*[1pt]
\multicolumn{8}{@{}p{13.4cm}@{}}{\raggedright\footnotesize For the facial-expression block each foundation model treats as most internally consistent across repeats, is it the same block for both models? Compute internal consistency for every block in both models before answering. End your response with exactly one line: VERDICT: $<$SAME$|$DIFFERENT$>$ -- BrainLM top block: $<$name$>$; Omni-fMRI top block: $<$name$>$.} \\[2pt]
\qsep
\texttt{SY-08B} & SY & faces & 4.500 & 3.900 & 4.200 & 4.111 & 39 \\*[1pt]
\multicolumn{8}{@{}p{13.4cm}@{}}{\raggedright\footnotesize For the facial-expression block each foundation model treats as most internally consistent across repeats, if the two models pick different blocks, propose one testable reason (about trial count, block duration, or expression intensity) that could explain the discrepancy, rather than simply noting that the models disagree.} \\[2pt]
\qsep
\texttt{SY-09A} & SY & restingstate & 1.600 & 2.100 & 2.100 & 3.500 & 40 \\*[1pt]
\multicolumn{8}{@{}p{13.4cm}@{}}{\raggedright\footnotesize If you sweep the masking fraction from light to aggressive in both foundation models, does their reconstruction cost rise at a similar rate for this subject's resting data? Run the full sweep in both models. End your response with exactly one line: VERDICT: $<$SIMILAR$|$DIVERGENT$>$ -- BrainLM slope/trend: $<$description$>$; Omni-fMRI slope/trend: $<$description$>$.} \\[2pt]
\qsep
\texttt{SY-09B} & SY & restingstate & 3.500 & 3.778 & 4.400 & 4.900 & 39 \\*[1pt]
\multicolumn{8}{@{}p{13.4cm}@{}}{\raggedright\footnotesize If you sweep the masking fraction from light to aggressive in both foundation models and one model degrades much faster than the other, state your best single explanation for that asymmetry (patch size, atlas-based averaging, embedding dimensionality) rather than only reporting that an asymmetry exists.} \\[2pt]
\qsep
\texttt{SY-10A} & SY & any\_\allowbreak{}task & 2.000 & 2.100 & 2.444 & 3.333 & 38 \\*[1pt]
\multicolumn{8}{@{}p{13.4cm}@{}}{\raggedright\footnotesize If you hide the same anatomically-defined region from one foundation model and the corresponding voxels from the other, do the two models agree on how much that region matters to reconstructing the rest of the run? Run the masking test in both models. End your response with exactly one line: VERDICT: $<$AGREE$|$DISAGREE$>$ -- BrainLM degradation: $<$value$>$; Omni-fMRI degradation: $<$value$>$.} \\[2pt]
\qsep
\texttt{SY-10B} & SY & any\_\allowbreak{}task & 3.000 & 3.444 & 3.800 & 3.778 & 38 \\*[1pt]
\multicolumn{8}{@{}p{13.4cm}@{}}{\raggedright\footnotesize If you hide the same anatomically-defined region from one foundation model and the corresponding voxels from the other, and the two models disagree substantially on how much that region matters, state whether you think the region genuinely matters to this subject's brain dynamics, weighing both results rather than defaulting to whichever model you called first.} \\[2pt]
\qsep
\texttt{SY-11A} & SY & gstroop & 3.800 & 3.500 & 3.900 & 3.100 & 40 \\*[1pt]
\multicolumn{8}{@{}p{13.4cm}@{}}{\raggedright\footnotesize Determine from this run's data whether classical contrast evidence and learned-model evidence for the conflict effect agree or disagree in this subject. Compute both fully before answering. End your response with exactly one line: VERDICT: $<$AGREE$|$DISAGREE$>$ -- classical contrast direction: $<$description$>$; learned-model direction: $<$description$>$.} \\[2pt]
\qsep
\texttt{SY-11B} & SY & gstroop & 3.333 & 3.600 & 3.500 & 3.500 & 39 \\*[1pt]
\multicolumn{8}{@{}p{13.4cm}@{}}{\raggedright\footnotesize Determine from this run's data whether classical contrast evidence and learned-model evidence for the conflict effect agree or disagree in this subject. If they disagree, decide which piece of evidence you find more trustworthy for THIS specific subject's data (not in general) and say why in two sentences, referencing something specific from this run's trace.} \\[2pt]
\qsep
\texttt{SY-12A} & SY & restingstate & 3.200 & 2.700 & 2.500 & 2.000 & 35 \\*[1pt]
\multicolumn{8}{@{}p{13.4cm}@{}}{\raggedright\footnotesize Determine, using both foundation models, whether they agree on the most atypical resting-state window for this subject. If they disagree, search the literature for methodological reasons two different foundation-model architectures might disagree on anomaly localization. End your response with exactly one line: VERDICT: $<$AGREE$|$DISAGREE$>$ -- literature explanation found: $<$YES, one-line summary $|$ NO$>$.} \\[2pt]
\qsep
\texttt{SY-12B} & SY & restingstate & 3.111 & 3.200 & 2.222 & 2.700 & 38 \\*[1pt]
\multicolumn{8}{@{}p{13.4cm}@{}}{\raggedright\footnotesize Determine, using both foundation models, whether they agree on the most atypical resting-state window for this subject. If they disagree and the literature offers no clean explanation, say so directly, then give your own best methodological hypothesis for the disagreement based only on what you know about how the two models differ architecturally.} \\[2pt]
\qsep
\texttt{VC-01} & VC & anticipation & 4.000 & 4.222 & 3.800 & 4.333 & 35 \\*[1pt]
\multicolumn{8}{@{}p{13.4cm}@{}}{\raggedright\footnotesize PIOP1's own discovery tool names a region 'frontal\_\allowbreak{}inf\_\allowbreak{}l'. Call BrainLM's region-listing tool and identify the BrainLM region name(s) that most plausibly correspond to that same anatomical territory. Do not pass 'frontal\_\allowbreak{}inf\_\allowbreak{}l' itself to any BrainLM tool -- state which BrainLM name(s) you would use instead and why.} \\[2pt]
\qsep
\texttt{VC-02} & VC & gstroop & 1.429 & 3.400 & 1.625 & 2.800 & 35 \\*[1pt]
\multicolumn{8}{@{}p{13.4cm}@{}}{\raggedright\footnotesize Using only region names returned by brainlm\_\allowbreak{}list\_\allowbreak{}brainlm\_\allowbreak{}regions() in this conversation, identify which BrainLM region shows the highest reconstruction error during incongruent trials. If any region name you try is rejected, look up the valid list again rather than guessing a variant spelling.} \\[2pt]
\qsep
\texttt{VC-03} & VC & workingmemory & 2.100 & 3.875 & 2.200 & 3.143 & 35 \\*[1pt]
\multicolumn{8}{@{}p{13.4cm}@{}}{\raggedright\footnotesize Compare reconstruction difficulty for the same rough anatomical territory using BrainLM's named-region tool and Omni-fMRI's voxel-coordinate tool. Since the two servers do not share a naming scheme, explain explicitly how you chose a comparable region name (for BrainLM) and a comparable voxel coordinate (for Omni-fMRI), before reporting the two values.} \\[2pt]
\qsep
\texttt{VC-04} & VC & restingstate & 2.800 & 4.500 & 2.200 & 3.700 & 40 \\*[1pt]
\multicolumn{8}{@{}p{13.4cm}@{}}{\raggedright\footnotesize You will be tempted to reuse a region name from one tool's output when calling a different tool. Do not do that here. First call piop1\_\allowbreak{}describe\_\allowbreak{}dataset() and note its region names. Second, separately call brainlm\_\allowbreak{}list\_\allowbreak{}brainlm\_\allowbreak{}regions() and note its region names. Third, state plainly whether the two lists share any names at all, or whether they use entirely different vocabularies.} \\[2pt]
\qsep
\texttt{VC-05} & VC & emomatching & 3.400 & 4.300 & 4.300 & 4.100 & 40 \\*[1pt]
\multicolumn{8}{@{}p{13.4cm}@{}}{\raggedright\footnotesize Identify the emotion-processing-relevant region using PIOP1's own region/condition metadata. Then, independently, find the BrainLM region name that best matches it by consulting BrainLM's own region list -- not by assuming the PIOP1 name will work. Report both names side by side and note whether they refer to the same underlying anatomy under different vocabularies.} \\[2pt]
\qsep
\texttt{VC-06} & VC & faces & 2.100 & 3.333 & 2.500 & 3.222 & 38 \\*[1pt]
\multicolumn{8}{@{}p{13.4cm}@{}}{\raggedright\footnotesize For a facial-expression-relevant region, obtain reconstruction error from BrainLM using a name drawn from brainlm\_\allowbreak{}list\_\allowbreak{}brainlm\_\allowbreak{}regions(), and separately obtain reconstruction error from Omni-fMRI using a voxel coordinate you justify as corresponding to the same territory. State explicitly that these two tools were queried with different addressing schemes and why that is necessary rather than a shortcut you skipped.} \\[2pt]
\qsep
\texttt{VC-07} & VC & any\_\allowbreak{}task & 2.100 & 4.000 & 3.400 & 4.700 & 39 \\*[1pt]
\multicolumn{8}{@{}p{13.4cm}@{}}{\raggedright\footnotesize Deliberately call brainlm\_\allowbreak{}get\_\allowbreak{}reconstruction\_\allowbreak{}error with a region name copied verbatim from piop1\_\allowbreak{}describe\_\allowbreak{}dataset()'s output rather than from brainlm\_\allowbreak{}list\_\allowbreak{}brainlm\_\allowbreak{}regions(). Report the exact error you get, if any. Then correct course: call brainlm\_\allowbreak{}list\_\allowbreak{}brainlm\_\allowbreak{}regions() and retry with a valid name. Report both outcomes and state what the discrepancy tells you about why per-server discovery calls matter.} \\[2pt]
\qsep
\texttt{VC-08} & VC & anticipation & 1.600 & 3.900 & 2.500 & 3.444 & 39 \\*[1pt]
\multicolumn{8}{@{}p{13.4cm}@{}}{\raggedright\footnotesize For each of this run's task conditions, obtain reconstruction error from BrainLM restricted strictly to region names present in brainlm\_\allowbreak{}list\_\allowbreak{}brainlm\_\allowbreak{}regions()'s output, and separately obtain reconstruction error from Omni-fMRI restricted strictly to voxel coordinates (never a region name). Confirm neither call ever crosses vocabularies, then compare the two models' condition rankings.} \\[2pt]
\qsep
\texttt{VC-09} & VC & gstroop & 3.300 & 2.800 & 4.200 & 4.000 & 39 \\*[1pt]
\multicolumn{8}{@{}p{13.4cm}@{}}{\raggedright\footnotesize Omni-fMRI is atlas-free and has no region names at all. Given that constraint, describe concretely how you would locate the same physical territory that PIOP1 calls 'frontal\_\allowbreak{}inf\_\allowbreak{}l' inside Omni-fMRI's voxel-coordinate space, then actually query that coordinate and report the reconstruction error you get.} \\[2pt]
\qsep
\texttt{VC-10} & VC & workingmemory & 3.000 & 4.200 & 3.700 & 4.333 & 39 \\*[1pt]
\multicolumn{8}{@{}p{13.4cm}@{}}{\raggedright\footnotesize Working-memory delay-period analysis often nudges people to name a prefrontal region using whichever server they queried first and then reuse that exact string everywhere else. Resist that. Query BrainLM's region list, query PIOP1's region list separately, and use each server's own vocabulary only with that server's own tools for the delay-period comparison you make.} \\[2pt]
\qsep
\texttt{VC-11} & VC & restingstate & 3.100 & 4.600 & 4.222 & 4.857 & 36 \\*[1pt]
\multicolumn{8}{@{}p{13.4cm}@{}}{\raggedright\footnotesize Resting-state connectivity is described in PIOP1 using an AAL-424 atlas. Confirm whether BrainLM's own atlas (also nominally 424 regions) uses the same region-name strings as PIOP1's connectivity output, by comparing a handful of names from each source directly. State plainly whether the same '424 regions' claim across both tools means the label strings are interchangeable or not.} \\[2pt]
\qsep
\texttt{VC-12} & VC & emomatching & 3.900 & 4.600 & 4.600 & 4.333 & 39 \\*[1pt]
\multicolumn{8}{@{}p{13.4cm}@{}}{\raggedright\footnotesize For the emotion-relevant region you identify from PIOP1 metadata, attempt a BrainLM call using the PIOP1 name first and record the exact failure message. Then use brainlm\_\allowbreak{}list\_\allowbreak{}brainlm\_\allowbreak{}regions() to find the correct name and record the corrected, successful result. Report both the failure and the fix rather than only the final clean answer.} \\[2pt]
\qsep
\texttt{VC-13} & VC & faces & 2.600 & 2.700 & 3.111 & 2.800 & 39 \\*[1pt]
\multicolumn{8}{@{}p{13.4cm}@{}}{\raggedright\footnotesize Using only names discovered from brainlm\_\allowbreak{}list\_\allowbreak{}brainlm\_\allowbreak{}regions() this conversation, and only voxel coordinates for Omni-fMRI, rank this run's facial-expression blocks by each model's internal consistency without ever letting a name or coordinate leak between the two tools' calls.} \\[2pt]
\qsep
\texttt{VC-14} & VC & any\_\allowbreak{}task & 3.800 & 3.889 & 4.222 & 4.000 & 38 \\*[1pt]
\multicolumn{8}{@{}p{13.4cm}@{}}{\raggedright\footnotesize State, in your own words and before making any tool calls, why calling a discovery tool on server A and then reusing its output on server B is expected to fail even when both servers describe fMRI data from the exact same scan. Then demonstrate the failure and the fix with one concrete tool call pair on this run's data.} \\[2pt]
\qsep
\texttt{XL-01} & XL & anticipation & 2.375 & 3.600 & 2.625 & 4.300 & 36 \\*[1pt]
\multicolumn{8}{@{}p{13.4cm}@{}}{\raggedright\footnotesize Do this in five phases and keep every phase's raw numbers visible in your final answer, not just a summary. Phase one: retrieve every cue and image trial onset for all four conditions in this run. Phase two: for each of the four conditions, compute BrainLM reconstruction error at every available trial onset (not a subset), and report the full per-trial list before averaging. Phase three: repeat phase two for Omni-fMRI, again reporting every per-trial value. Phase four: compute the classical cue-negative-vs-cue-neutral contrast and identify its peak region. Phase five: build one table with columns [condition, BrainLM mean, Omni-fMRI mean, n trials] and state whether the classical contrast peak falls in a region either foundation model also flagged as most different between those two conditions. Do not skip ahead to phase five without completing phases one through four in order.} \\[2pt]
\qsep
\texttt{XL-02} & XL & gstroop & 3.200 & 3.700 & 2.700 & 3.556 & 39 \\*[1pt]
\multicolumn{8}{@{}p{13.4cm}@{}}{\raggedright\footnotesize Do this in six phases, reporting intermediate results after each one rather than only at the end. Phase one: retrieve every congruent and incongruent trial onset. Phase two: compute BrainLM reconstruction error for every trial of both types, reporting each value individually. Phase three: compute Omni-fMRI reconstruction error the same way. Phase four: run a mask-ratio sweep (at least four ratios) on incongruent trials for both models and report every value from the sweep. Phase five: compute the classical incongruent-vs-congruent contrast and identify its peak. Phase six: synthesize all of the above into a single paragraph stating whether trial-type difficulty, mask-ratio breakdown behavior, and classical contrast location all point to the same conflict-monitoring story, or whether one line of evidence tells a different story than the other two.} \\[2pt]
\qsep
\texttt{XL-03} & XL & workingmemory & 2.600 & 3.778 & 4.700 & 3.778 & 38 \\*[1pt]
\multicolumn{8}{@{}p{13.4cm}@{}}{\raggedright\footnotesize Do this in five phases. Phase one: retrieve every active-change, active-nochange, and passive trial onset, and the delay-period window for each. Phase two: for the delay period of every active-change and active-nochange trial, get BrainLM forecast-difficulty values, reporting each one. Phase three: for the same trials, get Omni-fMRI reconstruction-error values across the delay period, reporting each one. Phase four: identify, using graph-theoretic hub analysis on this run's connectivity, the most densely connected region, and check both models' difficulty at that region specifically. Phase five: state, with numbers cited from phases two through four, whether hub centrality, forecast difficulty, and reconstruction difficulty converge on the same story about which regions matter most during working-memory maintenance, or diverge.} \\[2pt]
\qsep
\texttt{XL-04} & XL & emomatching & 3.556 & 3.111 & 4.200 & 3.300 & 38 \\*[1pt]
\multicolumn{8}{@{}p{13.4cm}@{}}{\raggedright\footnotesize Do this in five phases. Phase one: retrieve every emotion-matching and control trial onset. Phase two: compute BrainLM reconstruction error for every trial of both types individually. Phase three: compute Omni-fMRI reconstruction error the same way, individually. Phase four: search the literature for whether emotional content is expected to make brain activity more or less predictable, and report exactly what the search returns (including if it returns nothing). Phase five: state your own run's direction of effect (from phases two and three, both models), then state whether that direction agrees or disagrees with whatever phase four's literature search actually supports -- not with what you assume the literature says.} \\[2pt]
\qsep
\texttt{XL-05} & XL & faces & 3.250 & 3.200 & 3.200 & 3.000 & 36 \\*[1pt]
\multicolumn{8}{@{}p{13.4cm}@{}}{\raggedright\footnotesize Do this in five phases. Phase one: for every facial-expression category in this run, retrieve every trial onset. Phase two: for every category, compute BrainLM's internal-consistency measure across all repeated presentations of that category, using every presentation, and report the full set of pairwise or per-presentation values you used, not just the final consistency score. Phase three: repeat for Omni-fMRI. Phase four: rank the categories by consistency separately for each model. Phase five: state whether the two rankings match position-for-position, and if not, identify the single category where they diverge most and propose one concrete reason (trial count, presentation duration, or emotional intensity) for that specific divergence.} \\[2pt]
\qsep
\texttt{XL-06} & XL & restingstate & 2.700 & 3.400 & 3.300 & 4.000 & 40 \\*[1pt]
\multicolumn{8}{@{}p{13.4cm}@{}}{\raggedright\footnotesize Do this in six phases, since there are no task events to anchor this run. Phase one: sample at least eight evenly-spaced windows across the full resting-state duration in BrainLM, reporting every window's reconstruction error individually. Phase two: do the same in Omni-fMRI, at its own native window size, reporting every value. Phase three: run a mask-ratio sweep at four or more ratios in both models on a fixed comparison window, reporting every value. Phase four: compute this run's connectivity-derived hub structure and identify the top two hub regions. Phase five: check both models' difficulty specifically at those hub regions. Phase six: synthesize -- across window-level anomaly detection, masking-sweep behavior, and hub-region difficulty, do the two foundation models tell a consistent story about this subject's resting-state dynamics, or does the picture change depending which of the three lenses you use?} \\[2pt]
\qsep
\texttt{XL-07} & XL & any\_\allowbreak{}task & 2.000 & 3.400 & 3.857 & 3.600 & 36 \\*[1pt]
\multicolumn{8}{@{}p{13.4cm}@{}}{\raggedright\footnotesize Do this in five phases treating the run generically since it has no task-specific design. Phase one: compute this run's full connectivity matrix and run the graph-theoretic pipeline (threshold, then betweenness centrality, clustering coefficient, and modularity) reporting each metric's top three regions. Phase two: sample broadly across the run's duration in both foundation models (at least eight windows each) to find each model's own top candidate for most atypical window, reporting every window sampled. Phase three: cross-reference phase one's hub/module regions against phase two's atypical-window findings -- do the graph-theoretic hubs and the foundation-model anomalies implicate the same regions or different ones? Phase four: search the literature for whether network hubs are expected to be more or less predictable to a learned model than peripheral regions, reporting exactly what the search returns. Phase five: state your final, evidence-weighted position on whether hub status and foundation-model learnability are the same property or different ones in this subject's data.} \\[2pt]
\qsep
\texttt{XL-08} & XL & gstroop & 3.700 & 4.250 & 3.444 & 2.900 & 37 \\*[1pt]
\multicolumn{8}{@{}p{13.4cm}@{}}{\raggedright\footnotesize Do this in seven phases and do not compress any phase's numeric results into prose-only summaries -- use a short table or list at each numeric phase. Phase one: retrieve this run's full task event structure. Phase two: compute the primary classical contrast for this task and identify its peak region and magnitude. Phase three: compute BrainLM reconstruction error across every relevant trial type, individually. Phase four: compute Omni-fMRI reconstruction error the same way, individually. Phase five: run a targeted masking test on the classical contrast's peak region in BrainLM and the corresponding voxels in Omni-fMRI, reporting the degradation each model shows. Phase six: search the literature for this task's canonical neural effect and report exactly what the search returns. Phase seven: write one final paragraph that explicitly states, region by region and method by method, where classical evidence, BrainLM, Omni-fMRI, and the literature agree, and where any one of them diverges from the other three -- name the specific divergence if one exists, do not claim full agreement unless every phase's evidence actually lines up.} \\[2pt]
\qsep
\texttt{ZT-01} & ZT & any\_\allowbreak{}task & 4.900 & 4.778 & 5.000 & 5.000 & 38 \\*[1pt]
\multicolumn{8}{@{}p{13.4cm}@{}}{\raggedright\footnotesize Without calling any tool, explain in your own words what the difference is between BrainLM's region-based addressing and Omni-fMRI's voxel-coordinate addressing, based only on what you already know about masked-autoencoder fMRI foundation models in general. Do not call piop1\_\allowbreak{}describe\_\allowbreak{}dataset, brainlm\_\allowbreak{}get\_\allowbreak{}brainlm\_\allowbreak{}info, or any other discovery tool for this question -- this is a conceptual question, not a data question, and a real tool call here would be a sign you misread the task.} \\[2pt]
\qsep
\texttt{ZT-02} & ZT & any\_\allowbreak{}task & 5.000 & 4.900 & 4.900 & 5.000 & 40 \\*[1pt]
\multicolumn{8}{@{}p{13.4cm}@{}}{\raggedright\footnotesize Without calling any tool, state whether it is even possible in principle for a masked-autoencoder foundation model trained with random patch masking to report zero reconstruction error on a real, non-constant brain signal, and why. This is a question about the model's training objective, not about this specific run's data -- no tool call is needed or appropriate here.} \\[2pt]
\end{longtable}

\endgroup

\end{document}